# Nitrogen-containing Surface Ligands Lead to False Positives for Photofixation of $N_2$ on Metal Oxide Nanocrystals: An Experimental and Theoretical Study

*Daniel Maldonado-Lopez,[a,†] Po-Wei Huang,[b,†] Karla R. Sanchez-Lievanos,[c,†] Gourhari Jana,[a] Jose L. Mendoza-Cortes,[a,d,*] Kathryn E. Knowles,[c,*] and Marta C. Hatzell[b,e,*]*

[a] Department of Chemical Engineering & Materials Science, Michigan State University, East Lansing, MI 48824, USA

[b] School of Chemical & Biomolecular Engineering, Georgia Institute of Technology, Atlanta, GA 30318, USA

[c] Department of Chemistry, University of Rochester, Rochester, NY 14627, USA

[d] Department of Physics and Astrophysics, East Lansing, MI, 48824

[e] School of Mechanical Engineering, Georgia Institute of Technology, Atlanta, GA 30318, USA

† D.M-L., P.W.H., and K.R.S.L contributed equally to this work.

***Corresponding authors**: marta.hatzell@me.gatech.edu; kknowles@ur.rochester.edu; jmendoza@msu.edu

**Abstract**

Many ligands commonly used to prepare nanoparticle catalysts with precise nanoscale features contain nitrogen (e.g., oleylamine); here, we found that the use of nitrogen-containing ligands during the synthesis of metal oxide nanoparticle catalysts substantially impacted product analysis during photocatalytic studies. We confirmed these experimental results *via* hybrid Density Functional Theory computations of the materials' electronic properties to evaluate their viability as photocatalysts for nitrogen reduction. This nitrogen ligand contamination, and subsequent interference in photocatalytic studies, is avoidable through the careful design of synthetic pathways that exclude nitrogen-containing constituents. This result highlights the urgent need for careful evaluation of catalyst synthesis protocols, as contamination by nitrogen-containing ligands may go unnoticed since the presence of nitrogen is often not detected or probed.

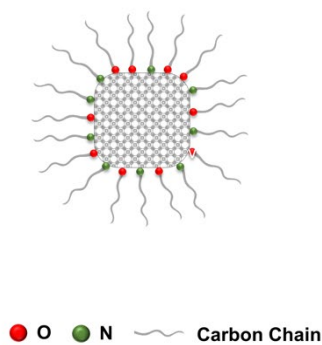
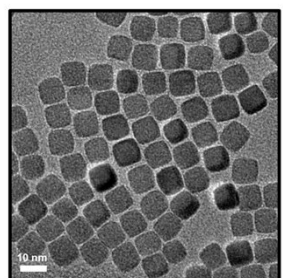
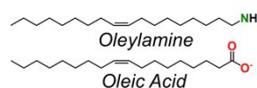
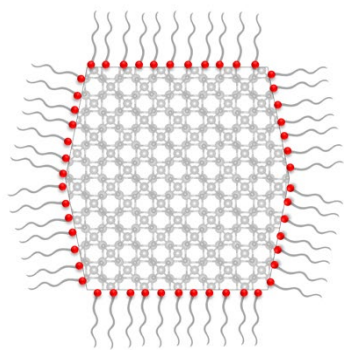
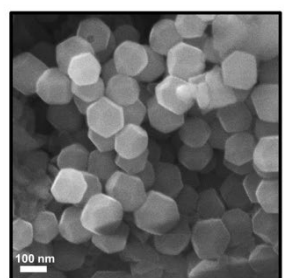
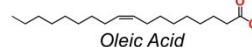

Ammonia is a vital fertilizer in agriculture, playing a crucial role in sustaining the needs of our growing population.[1] Current ammonia production is predominantly achieved *via* the Haber-Bosch process. However, this process is accompanied by substantial $CO_2$ emissions (accounting for 1.6% of global $CO_2$ emissions) and contributes to a worldwide nitrogen imbalance while consuming approximately 2% of the world's energy.[2] These pressing concerns require developing a decarbonized and energy-efficient approach to ammonia production. In this context, the photocatalytic nitrogen reduction reaction (pNRR) has garnered significant attention, as it enables catalysis under ambient conditions, utilizing water and air as readily available feedstocks.[3]

Recent studies suggest that a solar-to-ammonia (STA) efficiency as low as 0.1% but, more practically, 5-10% would be sufficient for the pNRR technique to meet fertilizer demand.[2] Despite notable progress in pNRR research, the best lab-scale studies currently achieve efficiencies that are at least an order of magnitude below the target, highlighting a significant gap that must be bridged before achieving commercial viability. Consequently, ongoing research endeavors are dedicated to enhancing the activity of photocatalysts, with a primary focus on developing and synthesizing nanoscale colloidal catalysts through meticulous materials design such as controlled shape, size, composition, and structure.[4,5] Researchers have developed several synthetic methods to access colloidally stable nanostructures of a broad range of materials, including quantum dots and complex metal oxides.[6] Notably, most of these methods use oleylamine ($C_{18}H_{35}NH_2$), a primary amine ligand that contains an amino group (-$NH_2$). These surface ligands play a crucial role in preventing catalyst aggregation, dictating morphology, and facilitating the formation of monodisperse nanocrystals. Hence, oleylamine and primary amine ligands have also surfaced in the synthesis of metal sulfide and oxyhalide catalysts for photo-(electro-)catalytic nitrogen fixation.[7,8]

Low ammonia yields in the current development of pNRR have also raised concerns about false-positive results.[9] These false-positive outcomes primarily stem from two factors: inaccurate low-level ammonia measurements or inadvertent ammonia contamination. Recent investigations have shed light on various sources of contamination, including the presence of ammonia and $NO_x$ species (such as NO, $N_2O$, and $NO_2$) in the feeding gas, surface contamination of equipment, and the utilization of N-containing catalysts.[10,11] In our experimental observations, we encountered severe ammonia contamination arising as a false positive from the presence of adventitious nitrogen, which is linked to nitrogen-containing ligands (NL) used during the catalyst synthesis process. We reinforce the appearance of false positives and explain the materials' catalytic behavior *via* hybrid Density Functional Theory (DFT). Our calculations suggest that the catalysts are inadequate for pNRR.

An ideal pNRR catalyst should be capable of absorbing visible light and exhibit both thermal and chemical stability. In this regard, spinel ferrite ($AFe_2O_4$) structures that exhibit a cubane-like motif (a ubiquitous cofactor in selective catalysts found in nature) are promising prospects for catalysis (see **Scheme 1**).[12]

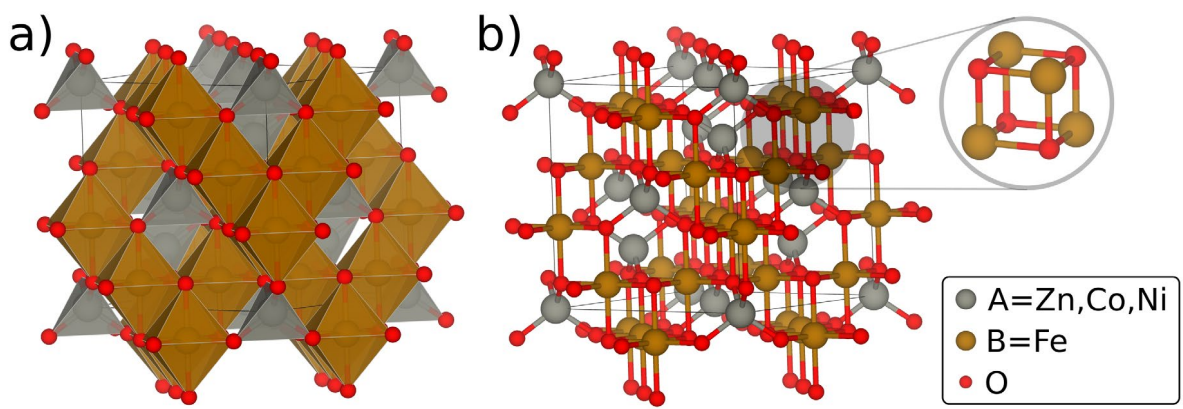

**Scheme 1**. Normal spinel ferrite structure. **a)** Polyhedral view. The octahedral and tetrahedral sites are shown in shaded brown and gray areas, respectively. **b)** Ball-and-stick representation, where the inset shows the cubane-like motif present in these structures.

In theory, the available versatility of the A site in spinel ferrites, along with the ability to vary the distribution of A and Fe among tetrahedral and octahedral crystal sites, can induce attractive synergy effects that may further improve catalytic activity. To explore this potential, we employed the synthetic method developed by Sanchez-Lievanos *et al.*, to synthesize several spinel ferrite nanomaterials, $AFe_2O_4$ (A: Co, Ni, and Zn).[13] We then evaluated their photocatalytic activity utilizing a custom-built photoreactor (**Fig. S1**). Metal ferrites were first dispersed in deionized (DI) water and then transferred into the photoreactor, which was then continuously aerated ($N_2$ or Ar) under irradiation (full spectrum). Following a 4-hour reaction period, we observed relatively high ammonia concentrations (1-2 ppm) for all three catalysts (**Fig. S2**). However, time-dependent measurements identified that most ammonia evolved before illumination. We also see that ammonia production quickly reaches a plateau after irradiation. This finding suggests that most of the ammonia we observed was from contaminants (nitrogen, amines, or ammonia-based) rather than being the product of a photocatalytic pathway. We meticulously examined all potential sources of ammonia contamination, quantified the experimental setup in its entirety, and demonstrated that—in the absence of the catalyst—background ammonia concentration was measured to be (0.0009 ± 0.0003 ppm), indicating that the catalyst itself was the source of contamination (**Fig. 1**).

While the metal ferrites employed in our study do not inherently contain nitrogen, amines, or ammonia, the ligands utilized during the nanocrystal synthesis (e.g., oleylamine) do possess these nitrogen-based constituents. To mitigate potential contamination from these amine groups, all catalysts were purified using standard procedures consisting of three precipitation cycles with a mixture of ethanol and hexane followed by centrifugation. However, contamination persisted even in catalysts that had undergone what is typically considered an effective oleylamine removal process. These "purified" catalysts were utilized in our experiments and are denoted as NL in **Fig. 1**. The presence of this organic molecule on the catalyst surface is postulated as the primary cause of the observed contamination. These findings indicate that standard washing treatments do not effectively eliminate all ligands.

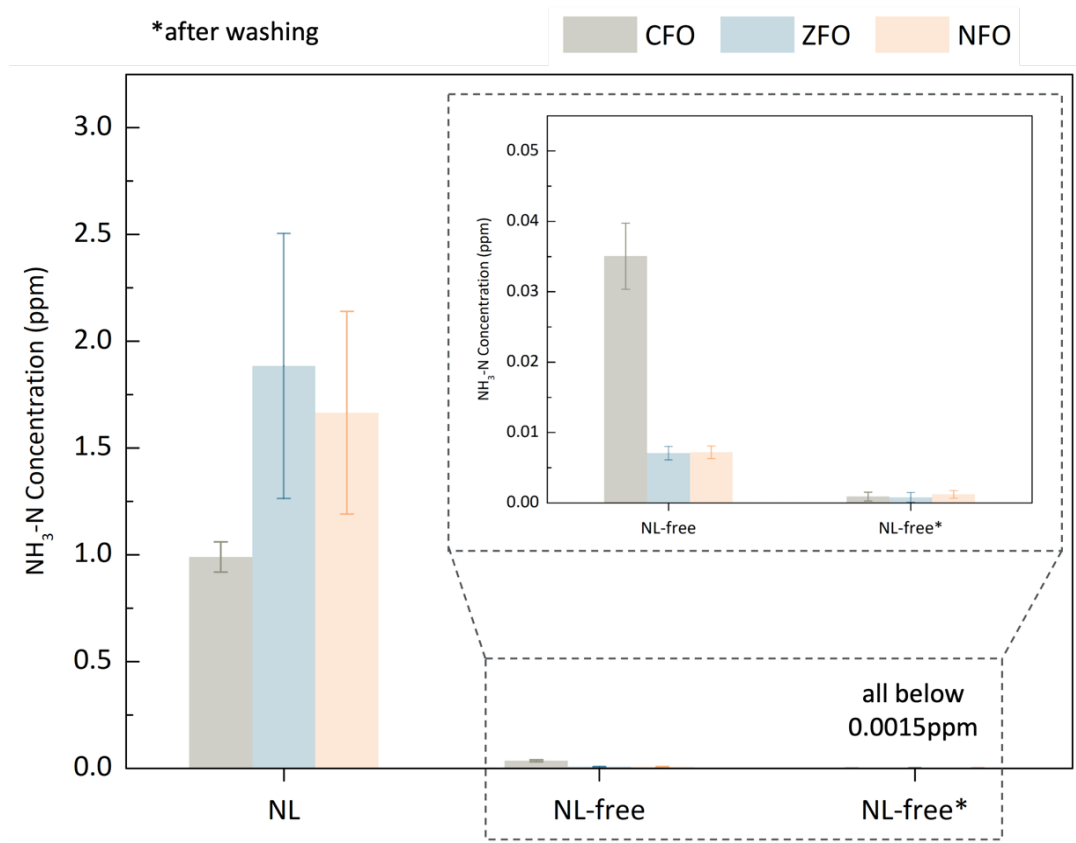

**Figure 1.** Ammonia detected by ion chromatography in solutions of Co (CFO), Zn (ZFO), and Ni (NFO) ferrite prepared with and without nitrogen-containing ligands (NL). The sample was collected after 30 minutes of sonication (catalyst in DI water) in the absence of illumination. The error bars represent the standard deviation from three independent experiments.

Next, we redesigned the synthetic route, entirely eliminating the use of nitrogen-containing ligands to produce the metal ferrites. These catalysts are labeled as NL-free in **Fig. 1**. The resulting NL-free nanocrystals are significantly larger than the NL samples, with diameters on the order of 100s of nm instead of ~15 nm obtained using oleylamine (Figures S8 and S9). Comparison of the compositions of the NL and NL-free nanocrystals by energy-dispersive x-ray spectroscopy (X-ray) is consistent with successful elimination of nitrogen from the NL-free nanocrystals (**Fig. S11**). As a result of this modification, ammonia contamination in the NL-free ferrites was significantly reduced. Furthermore, we observed that residual contamination could be further suppressed by rinsing the catalysts with deionized (DI) water. This rinsing procedure involved three cycles of washing with DI water, followed by centrifugation. These catalysts are labeled as NL-free* (**Fig. 1**). Our results provide compelling support for our hypothesis that the presence of nitrogen ligands can introduce substantial ammonia contamination. Illumination of the NL-free* nanocrystals under $N_2$ and Ar atmospheres reveals statistically significant photoinduced production of ammonia from nitrogen for $NiFe_2O_4$ (NFO) only; $CoFe_2O_4$ (CFO) and $ZnFe_2O_4$ (ZFO) produced similar concentrations of ammonia under illumination in both $N_2$ and Ar atmospheres (**Fig. 2**).

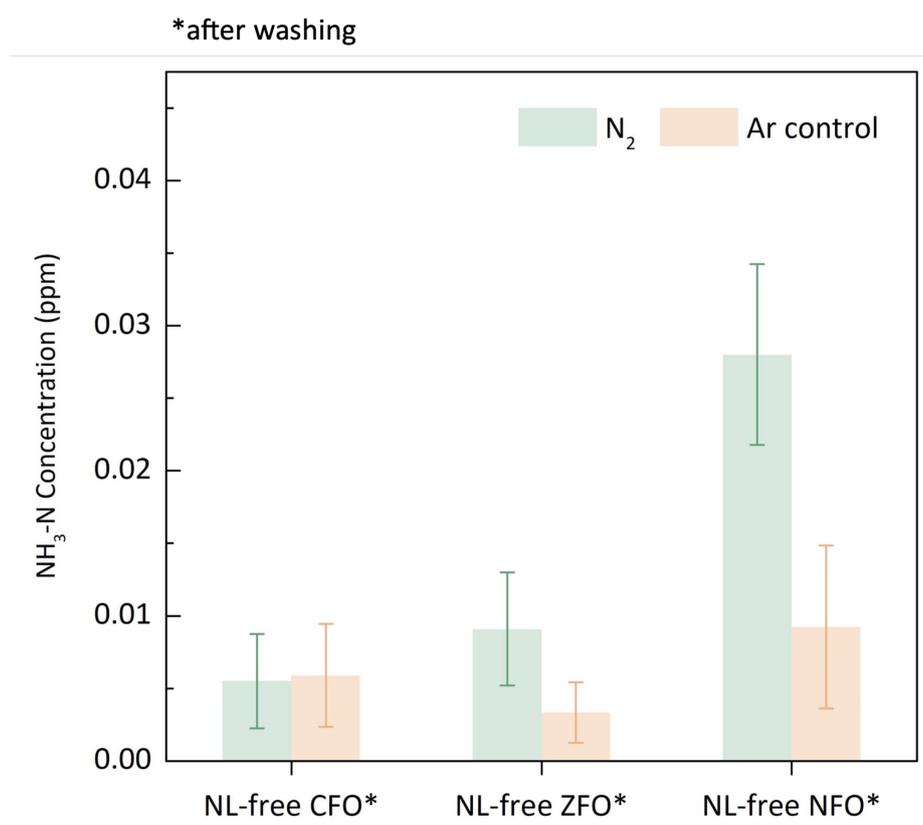

**Figure 2.** Photocatalytic activity of post-washed NL-free Co, Zn, and Ni ferrites under full-spectrum irradiation after 4 hours in $N_2$ or Ar environments. The error bars represent the standard deviation from two independent experiments.

To support these experimental results, we performed an in-depth computational analysis of the ZFO, CFO, and NFO structures using hybrid-level DFT. The computational methodology is described in detail in the supplementary information (SI). These structures have a general formula $(A_{1-x}B_x)[A_xB_{2-x}]O_4$, where the parentheses describe tetrahedral sites and the square brackets represent octahedral sites. In our spinel ferrites, A = Zn, Co, Ni, and B = Fe. The inversion parameter, $x$, describes the fraction of tetrahedral sites occupied by $Fe^{3+}$ ions and the fraction of $A^{2+}$ ions that occupy octahedral sites. Based on X-ray photoelectron spectroscopy, it was determined that the NL-free nanoparticles have inversion parameters $x \sim 0$ for ZFO and $x = 0.8$ for CFO, which closely resemble the inversion parameters of the bulk materials (see **Fig. S12**).[14-16] Since NFO strongly prefers an inverted configuration, we assume $x = 1$ for NFO.[17] The bulk structures are shown in **Fig. S13**, and the corresponding crystallographic data is available in the SI. Moreover, scanning electron microscopy imaging (**Fig. S9**) indicates that ZFO NPs are primarily octahedral in morphology, indicating the predominance of (111) surface planes. On the other hand, CFO and NFO NPs most closely resemble truncated octahedra, which would indicate the presence of both (111) and (100) planes. Therefore, slab calculations were set up for these surface planes with inversion parameters of $x = 0$ for ZFO, $x = 0.75$ for CFO, and $x = 1$ for NFO.

Due to the high reactivity of exposed transition metals in the (100) slabs, hydrogen atoms were added on both sides of the surface reconstructions, effectively "capping" the surface. Similarly, due to the (111) slabs' termination, the addition of oxygen and hydrogen atoms was necessary to avoid exposed transition metals. In particular, hydrogen atoms were added to either the top or bottom surfaces to maintain a charge balance, resulting in two configurations. Although the inclusion of oxygen and hydrogen alters the compounds' stoichiometry, these structures are more reminiscent of the surfaces that would be observed in an experimental setting due to exposing the NPs to ambient $O_2$ and $H_2O$. Side views of the optimized surface reconstructions are shown in **Fig. 3** and crystallographic data is available in the SI.

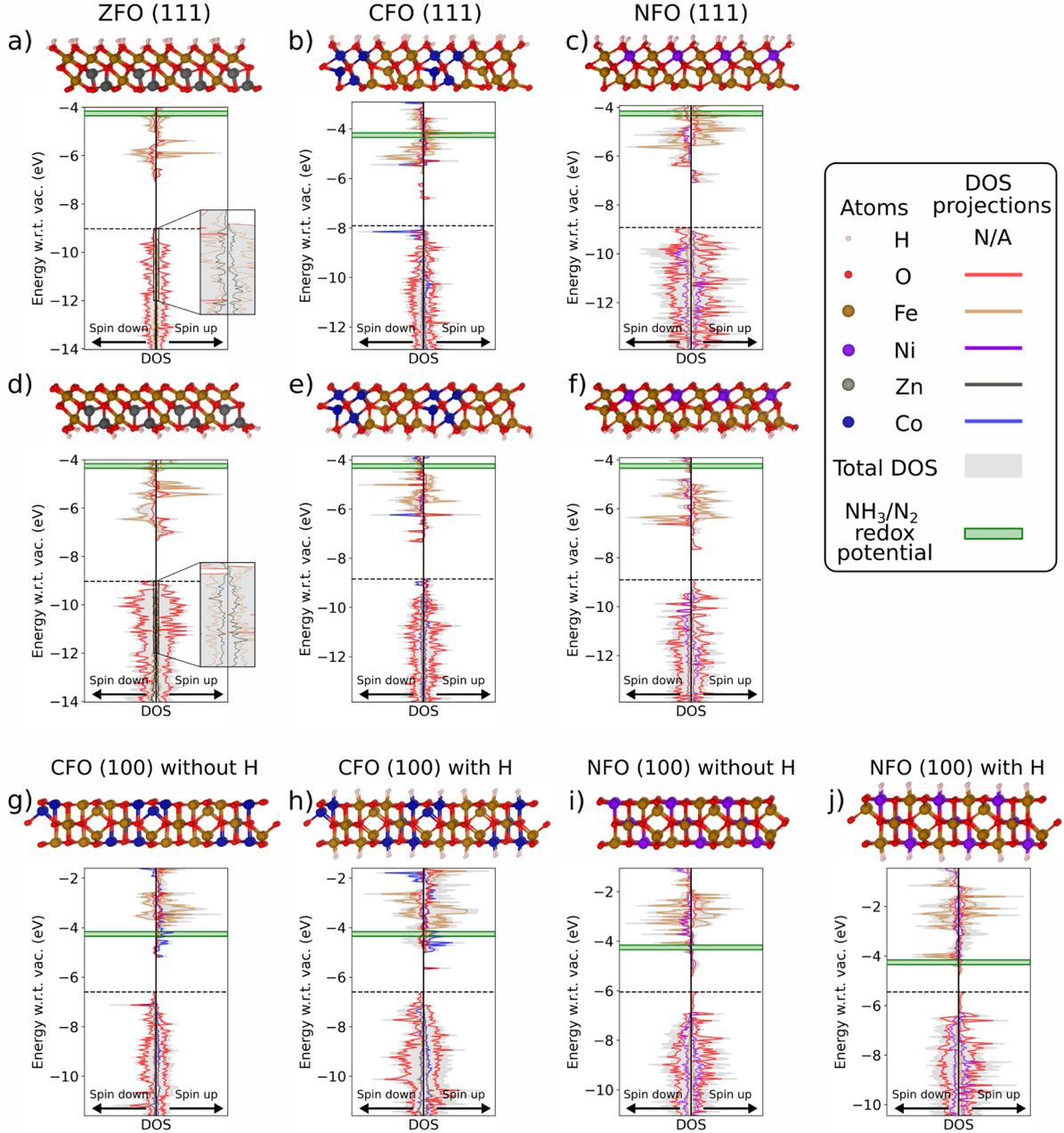

**Figure 3.** Spin-polarized atom-projected electronic density of states of ZFO, CFO, and NFO slabs, and side views of the corresponding structures. **a-c)** correspond to (111) slabs with H added to the top layer, **d-f)** correspond to (111) slabs with H added to the bottom layer, and **g-j)** show (100) slabs without H and with H added to both exposed surfaces. The left and right-hand sides of each plot represent spin-down and spin-up electronic channels, respectively. Total DOS is plotted as a shaded gray area, while the contribution from each atomic species in the material was projected and shown as lines. The DOS plots are aligned with respect to vacuum, and the $NH_3/N_2$ potential is plotted as a green rectangle. The DOS contributions from H are not shown, as they did not present a significant contribution to the total DOS.

**Fig. 3** plots the spin-polarized atom-projected electronic density of states (DOS) for the ZFO (111) slab (with *x=0*, i.e. NL-free), CFO (111) and (100) slabs, and NFO (111) and (100) slabs. For the (111) slabs, we show the hydrogen capped structures. Structures without hydrogen are plotted in **Fig. S14** and show similar results. For the (100) slabs, we plot the structures with and without hydrogen capping. Furthermore, band alignments to an absolute vacuum scale were performed in these calculations, allowing us to observe the relative electron density at different electrode potentials. For this work, we are interested in the $NH_3/N_2$ potential, which falls at around -0.28 to -0.092 V vs. the Normal Hydrogen Electrode (NHE),[18-21] or -4.35 to -4.16 eV with respect to vacuum (the range being attributed to different pH conditions). Generated photoelectrons must have energy above this threshold, ideally *with the conduction band minimum (CBM) above the $NH_3/N_2$ potential*, for a material to be effective as a photo-(electro-)catalyst for ammonia generation.

We begin by analyzing the (111) surfaces (**Fig. 3 a-f**). Structurally, the slabs are formed by one layer with octahedral sites only (top layer) and a second layer with octahedral and tetrahedral sites (bottom layer). From their densities of states, it can be observed that the (111) slabs are spin-polarized, i.e., they are asymmetric over their spin-up and spin-down channels, which is expected from their magnetic configuration. Moreover, the addition of hydrogen to the top or bottom surface slightly changes the levels that form around the Fermi level. This is reflected in the slabs' band gaps, which are shown in **Table 1** and range from 1.1 to 2.7 eV. From the DOS projections, we observe that ZFO presents a large contribution from Fe and O atoms around the Fermi level, and a smaller contribution from Zn (see insets in **Fig. 3a** and **3d**). Comparatively, CFO (**Fig. 3b** and **3e**) has relatively large contributions from Co near the valence band maximum (VBM), especially when H atoms are added to the top (octahedral) layer. Furthermore, the band gap of CFO is smaller when adding H to the top surface due to the presence of inter-gap states, which are not as prevalent when adding H to the bottom layer. NFO, on the other hand, presents contributions from Ni throughout the valence band and has very similar DOS profiles when adding H to the bottom and top surfaces.

**Table 1**. Spin-up and spin-down band gaps for (111) slabs.

| Structure | Band gap (eV) | |
|---|---|---|
| | Spin-up channel | Spin-down channel |
| ZFO (111) H top | 2.53 | 1.97 |
| CFO (111) H top | 1.30 | 1.11 |
| NFO (111) H top | 1.99 | 2.51 |
| ZFO (111) H bottom | 1.83 | 2.25 |
| CFO (111) H bottom | 1.71 | 1.64 |
| NFO (111) H bottom | 1.20 | 2.74 |

A crucial result from the (111) slab calculations is that their CBM falls well below the $NH_3/N_2$ energy range, which is not conducive to ammonia production. This result is also consistent with the slabs without hydrogen (see **Fig. S14**). In the synthesized ZFO nanoparticles, the (111) slab is the dominating exposed surface; therefore, this analysis discards the ZFO structure as a potential candidate for nitrogen activation and reinforces our experimental findings, where ZFO NPs did not produce photoinduced ammonia. Note that this analysis corresponds to the NL-free ZFO structure ($x=0$); however, similar results are observed when comparing to the NL structure with inversion $x=⅔$ (see **Fig. S15**).

Next, we focus on the CFO (100) slabs (**Fig. 3g-h**). These structures are highly spin-polarized, and their spin-up and spin-down CBM/VBM fall at different relative energies, resembling the electronic character of a bipolar magnetic semiconductor. The structure with hydrogen capping presents intermediate states similar to the (111) slab, which arise from Co and O contributions. These states are not present in the structure without hydrogen. The CFO (100) slabs have bandgaps of 1.87 eV (spin-up) and 1.52 eV (spin-down) for the structure without H, while the structure with H has bandgap values of 1.50 eV and 1.75 eV for its spin-up and spin-down channels, respectively. In terms of their potential for ammonia generation, it can be observed that both slabs have CBM below the energy range of interest, preventing photoelectrons with sufficient energy from being generated to catalyze the $NH_3/N_2$ reaction. Therefore, the CFO structure is not a candidate for ammonia generation, in alignment with our experimental results.

Finally, we analyze the NFO (100) slabs (**Fig. 3i-j**). Our first observation is that the structures are highly spin-polarized, with a wide band gap in one of their spin channels and a narrow band gap in the other, resembling the electronic structure of a half semiconductor. They present bandgaps of 0.62 eV (spin-up) and 2.45 eV (spin-down) for the structure without H, and 0.69 eV (spin-up) and 2.0 eV (spin-down) for the structure with H. The slabs' CBM and VBM for their spin-up channels are mainly composed of Fe and O atomic orbitals. Meanwhile, their spin-down channel CBMs are mainly composed of Fe, and their VBMs are composed of O, Ni, and Fe contributions (in that order, from greater to lesser contribution). It can be observed that their spin-up CBM falls well below the $NH_3/N_2$ potential. However, their spin-down CBM is very close to the energy range of interest, therefore, we believe that photoexcitation could allow some electrons to undergo the transition from VBM to CBM in the spin-down channel, resulting in carriers with enough energy to catalyze the reaction. As previously discussed, the (100) surface is only partially exposed in the nanoparticles, and the reaction can only occur with electrons excited in the spin-down channel. Therefore, this mechanism would explain why we obtain a statistically significant but relatively small amount of photoinduced ammonia from NFO nanocrystals. In general, our electronic property calculations show a similar behavior as the experimental results for NL-free NPs. This reinforces our observation of the appearance of false-positives when utilizing nitrogen-containing species during nanocrystal synthesis.

In conclusion, it is crucial to recognize nitrogen-containing ligands, specifically oleylamine, as an additional source of ammonia contamination in photo-(electro-)catalytic ammonia synthesis. To mitigate the risk of surface-bonded amine contamination, we recommend adopting a nitrogen-free synthesis platform. Moreover, it is imperative to thoroughly discuss the synthesis methods employed in generating nanoparticle catalysts, as the presence of nitrogen-containing ligands may not be immediately apparent from the chemical formula of the catalyst itself. Neglecting the careful handling of nitrogen-containing ligands can result in ammonia measurements from contamination that exceed the actual photocatalytic activity by more than two orders of magnitude, severely impacting the identification of photo-fixed ammonia. The attention towards nitrogen ligand contamination should extend beyond the field of pNRR and encompass related areas, such as photo-(electro-)catalytic urea synthesis, which employs the co-reduction of nitrogen and carbon dioxide to ammonia and carbon monoxide as precursors for urea production.[22] Reducing adventitious ammonia requires re-evaluating material activity to ensure reliable results and interpretations.

**Supporting Information.** Details of the nanocrystal synthesis, photocatalysis experiments, ammonia quantification protocol, and computational methods; nanocrystal characterization data including electron microscopy images (TEM and SEM), energy dispersive x-ray spectroscopy (EDS), powder x-ray diffraction (pXRD) and x-ray photoelectron spectroscopy (XPS); photocatalysis control experiments; crystallographic information for all slab calculations, computed bulk electronic structures, and additional supporting slab calculations.

**Acknowledgements.** This work was supported financially by a Scialog Negative Emissions Sciences grant funded by the Alfred P. Sloan foundation. K. R. S.-L. was also supported in part by the National Science Foundation under grant (CHE-2044462). This work was supported in part through computational resources and services provided by the Institute for Cyber-Enabled Research at Michigan State University.**References**

1. Brightling, J. Ammonia and the Fertiliser Industry: The Development of Ammonia at Billingham. *Johnson Matthey Technology Review* **2018**, *62* (1), 32–47.

2. Medford, A. J.; Hatzell, M. C. Photon-Driven Nitrogen Fixation: Current Progress, Thermodynamic Considerations, and Future Outlook. *ACS Catal.* **2017**, *7* (4), 2624–2643.

3. Zhao, Y.; Miao, Y.; Zhou, C.; Zhang, T. Artificial Photocatalytic Nitrogen Fixation: Where Are We Now? Where Is Its Future? *Mol. Catal.* **2022**, *518*, 112107.

4. Shi, R.; Zhao, Y.; Waterhouse, G. I.; Zhang, S.; Zhang, T. Defect engineering in photocatalytic nitrogen fixation. *ACS Catal.* **2019**, *9*(11), 9739-9750.

5. Chen, C.; Liang, C.; Xu, J.; Wei, J.; Li, X.; Zheng, Y.; Li, J.; Tang, H.; Li, J. Size-Dependent Electrochemical Nitrogen Reduction Catalyzed by Monodisperse Au Nanoparticles. *Electrochim. Acta* **2020**, *335*, 135708.

6. Mourdikoudis, S.; Liz-Marzán, L. M. Oleylamine in Nanoparticle Synthesis. *Chem. Mater.* **2013**, *25* (9), 1465–1476.

7. Li, P.; Gao, S.; Liu, Q.; Ding, P.; Wu, Y.; Wang, C.; Yu, S.; Liu, W.; Wang, Q.; Chen, S. Recent Progress of the Design and Engineering of Bismuth Oxyhalides for Photocatalytic Nitrogen Fixation. *Adv. Energy and Sustainability Res.* **2021**, *2* (5), 2000097.

8. Chen, X.; Ma, C.; Tan, Z.; Wang, X.; Qian, X.; Zhang, X.; Tian, J.; Yan, S.; Shao, M. One-Dimensional Screw-like $MoS_2$ with Oxygen Partially Replacing Sulfur as an Electrocatalyst for the $N_2$ Reduction Reaction. *Chem. Eng. J.* **2022**, *433*, 134504.

9. Iriawan, H.; Andersen, S. Z.; Zhang, X.; Comer, B. M.; Barrio, J.; Chen, P.; Medford, A. J.; Stephens, I. E. L.; Chorkendorff, I.; Shao-Horn, Y. Methods for Nitrogen Activation by Reduction and Oxidation. *Nat. Rev. Methods Primers* **2021**, *1* (1), 56.

10. Choi, J.; Suryanto, B. H. R.; Wang, D.; Du, H.-L.; Hodgetts, R. Y.; Ferrero Vallana, F. M.; MacFarlane, D. R.; Simonov, A. N. Identification and Elimination of False Positives in Electrochemical Nitrogen Reduction Studies. *Nat. Commun.* **2020**, *11* (1), 5546.

11. Huang, P.-W.; Hatzell, M. C. Prospects and Good Experimental Practices for Photocatalytic Ammonia Synthesis. *Nat. Commun.* **2022**, *13* (1), 7908.
12

# Supporting Information

# Nitrogen-containing Surface Ligands Lead to False Positives for Photofixation of $N_2$ on Metal Oxide Nanocrystals: An Experimental and Theoretical Study

*Daniel Maldonado-Lopez,[a,†] Po-Wei Huang,[b,†] Karla R. Sanchez-Lievanos,[c,†] Gourhari Jana,[a] Jose L. Mendoza-Cortes,[a,d,*] Kathryn E. Knowles,[c,*] and Marta C. Hatzell[b,e,*]*

[a] Department of Chemical Engineering & Materials Science, Michigan State University, East Lansing, MI 48824, USA

[b] School of Chemical & Biomolecular Engineering, Georgia Institute of Technology, Atlanta, GA 30318, USA

[c] Department of Chemistry, University of Rochester, Rochester, NY 14627, USA

[d] Department of Physics and Astrophysics, East Lansing, MI, 48824

[e] School of Mechanical Engineering, Georgia Institute of Technology, Atlanta, GA 30318, USA

† D.M-L., P.H., and K.R.S.L contributed equally to this work.

*Corresponding authors: marta.hatzell@me.gatech.edu; kknowles@ur.rochester.edu; jmendoza@msu.edu

*Materials:*

**Synthesis of nitrogen-ligand (NL) containing MFe$_2$O$_4$ nanocrystals from single-source precursors under solvothermal conditions.**

In general, the single-source precursor, a trinuclear oxo-centered cluster with formula MFe$_2$(μ$_3$-O)(μ$_2$-O$_2$CCF$_3$)$_6$(H$_2$O)$_3$ (0.25 mmol, synthesis procedure adopted from ref 1),[1] oleylamine (OAm (≥98%), 27 mmol, Sigma-Aldrich), oleic acid (OA (90%), 27 mmol, Sigma-Aldrich), and benzyl ether (BE (99%), 40 mL, Sigma-Aldrich) were added to a 100 mL Teflon insert. The mixture was stirred for 15 minutes under ambient conditions to form a clear dark suspension. Subsequently, the Teflon insert was loaded into a stainless-steel autoclave, sealed, and heated at 230 °C for 24 h. The autoclave was allowed to cool down over a period of 8-12 hours under a well-ventilated fume hood. The suspension was then purified with addition of a 10:1 mixture of ethanol:hexane, followed by centrifugation. This washing step was conducted a total of three times.

**Synthesis of NL-free MFe$_2$O$_4$ nanoparticles from stoichiometric mixtures of metal acetylacetonate salts (acac) under solvothermal conditions.**

MFe$_2$O$_4$ nanoparticles were synthesized using stoichiometric (2:1) mixtures of Fe(acac)$_3$ (0.17 mmol, 97%, Sigma-Aldrich), M(acac)$_2$ (0.08 mmol, M(II): Ni (95%), Co (97%) and Zn (≥ 95.0 %), Sigma-Aldrich), dissolved in 40 mL of benzyl ether (BE (99%), 40 mL, Sigma-Aldrich) with oleic acid (OA, 27 mmol) and heated to 230 ºC for 24 h in a 100-mL autoclave reactor. The autoclave was allowed to cool down over a period of 8-12 hours under a well-ventilated fume hood. The suspension was then purified with addition of a 10:1 mixture of ethanol:hexane, followed by centrifugation. This washing step was conducted a total of three times.

*Characterization Methods:*

**Powder X-ray diffraction.**

The use of a copper source for powder X-ray diffraction analysis of iron-containing samples yields a high background signal due to X-ray fluorescence from iron. Monochromators can be used to suppress this background fluorescence signal; however, this approach presents disadvantages, such as low penetration depth and loss of peak intensity, that can lead to diffractograms with inconclusive peak patterns. Employing molybdenum radiation can help overcome these drawbacks. The lower fluorescence background signal observed using this radiation provides an improved signal-to-noise ratio and enables unambiguous phase identification. We therefore performed powder X-ray diffraction measurements using a Rigaku XtaLAB Dualflex Synergy-S diffraction system with Mo K$\alpha$ radiation ($\lambda$ = 0.71073 Å). We used the Bragg equation to convert the 2θ values obtained using the Mo source to 2θ values corresponding to the wavelength of a Cu K$\alpha$ source ($\lambda$=1.54148 Å) to compare our measured spectra to standard data deposited in the JCPDS database that was collected with Cu K$a$ radiation. Samples for powder X-ray diffraction measurements were prepared by drop casting hexane dispersions of purified nanocrystals onto glass substrates under ambient atmosphere. Prior to film fabrication, the glass substrates were cleaned with 3 cycles of hexane, water and isopropanol each under sonication for 15 minutes per cycle.

**Scanning Electron Microscopy (SEM).**

SEM images were taken using a Zeiss Auriga Scanning Electron Microscope with an electron beam energy of 20 kV. The nanocrystal samples were dropcast onto cleaned Si wafers from dispersions in hexane and grounded to instrument-specific aluminum specimen stubs using carbon tape. The diameter of the particles was measured using ImageJ software.

**Transmission Electron Microscopy (TEM)**

TEM micrographs and selected area electron diffraction (SAED) patterns were obtained using a FEI Tecnai F20 TEM with a beam energy of 200kV. The nanocrystal samples were drop-casted onto lacey carbon copper grids from hexane dispersions. The diameter of the particles was measured using ImageJ software

**Energy Dispersive X-Ray Spectroscopy (EDS).**

Energy dispersive X-ray spectroscopy elemental maps were obtained using a Zeiss Auriga Scanning Electron Microscope with an EDS analyzer and EDAX Apex software. Measurements were carried out using a 25 kV electron beam.

**X-ray Photoelectron Spectroscopy (XPS).**

XPS sample preparation was performed under ambient atmosphere. The nanocrystal powders were dispersed in hexane to obtain a concentrated solution. The solution was drop-casted onto cleaned Si wafers, which were electrically grounded to the sample bar by carbon tape. The XPS measurements were recorded with a Kratos Axis Ultra DLD system equipped with a monochromatic Al Kα (hv=1486.6eV) X-ray source. During the measurements, pressure in the main chamber was kept below $1\times10^{-7}$ mbar. Charge compensation was carried out via a neutralizer running at a current of $7 \times 10^{-6}$ A, a charge balance of 5 eV, and a filament bias of 1.3 V. The X-ray gun was set to 10 mA emission. Binding energies were referenced to the C 1s peak arising from adventitious carbon with binding energy of 284.8 eV. The C 1s, O 1s, Fe 2p and M 2p core levels were recorded with a pass energy of 80 eV. We collected three scans for iron, M, and oxygen, and two scans for carbon. XPS analysis was performed with CasaXPS (Version 2.3.22PR1.0.) The U Touggard function was used for background subtraction and the peaks were fitted with one or more Gaussian components. The XPS signals were fitted with the CasaXPS Component Fitting tool.

*Photoactivity Test of Metal Ferrites:*

**Photocatalytic reaction experiment.**

The photocatalytic performance of the metal ferrites was evaluated under full-spectrum irradiation, using a 300 W Xenon lamp (Newport Corporation) as the light source. In a typical experiment, 30 mg of metal ferrite was added to 30 mL of deionized water, followed by 30 min of sonication to obtain a well-dispersed mixture. The mixture was then transferred to a homemade photoreactor (Fig. S1). The photoreactor was equipped with a water recirculationattachment to maintain a constant temperature of 23 °C. The mixture was continuously stirred in the dark (200 rpm) whilst ultra-high-purity $N_2$ or Ar (Airgas) was bubbled through the solution at a flow rate of 50 mL/min for 30 min to obtain a saturated aqueous solution. To avoid contamination from the feed gases ($N_2$ and Ar), the gases were pretreated through an alkaline trap (0.1 M $KMnO_4$ in 0.1 M KOH) and a water trap to remove adventitious ammonia and $NO_x$.[2] The photoreactor was then continuously aerated with $N_2$ or Ar at a 50 mL/min flow rate under full-spectrum irradiation with continuous stirring. All the equipment, vial, tube, filter, and caps were cleaned with DI water to suppress ammonia contamination before every experiment (and measurement). The NL-free metal ferrites used for the photocatalytic reaction experiment all undergo a standard washing procedure (labeled as NL-free*). The washing procedure is to mix metal ferrite with deionized water at a ratio of 1 mg to 1 ml, then perform ultrasonic treatment (5 minutes), centrifuge, and repeat this process three times.

**Ammonia quantification.**

5 mL of the reaction mixture was collected at the corresponding time and passed through a PTFE syringe filter (Foxx Life Sciences) to obtain a transparent solution for the ammonia measurement. The concentration of ammonia in the solution was determined by Dionex Aquion ion chromatography (Thermofisher) coupled with CS12A cation exchange column, 20 mM methanesulfonic acid (Sigma-Aldrich) was used as the eluent passing through the column with a flow rate of 0.25 mL/min. A calibration curve (Fig S3) was made with ammonia standard solution (Hach Company). A limit of detection of 0.0003 ppm was obtained using this ion chromatography method.

*Computational Methods:*

**Computational Parameters** Calculations were carried out using unrestricted Density Functional Theory within the CRYSTAL17 software suite,[3] which employs Gaussian-type orbitals (GTOs), allowing an efficient implementation of post-HF methods, such as hybrid functionals. The exchange-correlation functional was defined through the hybrid Heyd-Scuseria-Ernzerhof 2006 (HSE06)[4] formulation, which was used for all geometry optimization and single-point energy computations. Long-range interactions were taken into consideration via Grimme's third-order (-D3) dispersion corrections.[5] Triple-zeta with polarization (TZVP) quality basis sets were employed during all calculations.[6]

The k-points used to compute the Hamiltonian matrix were sampled using a Pack-Monkhorst grid with a resolution approximately between $\frac{1}{b_i}\frac{2\pi}{40}$ and $\frac{1}{b_i}\frac{2\pi}{60}$, where $b_i$ are reciprocal lattice vectors. On the other hand, the density matrix and Fermi energy were computed on a denser Gilat k-point grid, with a resolution between $\frac{1}{b_i}\frac{2\pi}{80}$ and $\frac{1}{b_i}\frac{2\pi}{120}$.

**Bulk Spinel Structural Ground State**

Inverted spinels are intrinsically disordered systems; therefore, a combinatorial analysis must be performed to find their ground state. The cell size and number of inverted sites depend on their inversion parameter. To find the most stable structures, bulk spinel (($A_{1-x}Fe_x$)[$A_xFe_{2-x}$]$O_4$ general formula, with A = Zn, Co, and Ni, and corresponding inversion parameters of 2/3, 0.75, and 1) calculations were run with different atomic arrangements. The process was carried out as follows. First, the normal spinel structures (x=0) were fully optimized using their primitive unit cell, which has 14 atoms (two tetrahedral sites, four octahedral sites and eight oxygen sites). During this optimization, an initial Néel-type ferrimagnetic order was set, but the unrestricted wave functions were allowed to drift to any magnetic structure that minimized their energy. Next, the inverted structures were built from the optimized normal spinel structures. This was achieved by exchanging A and B sites. Note that for x=0.75 and x=0.67 it was necessary to build 1✕1✕2 and 1✕1✕3 supercells to have an adequate number of A and Fe sites to achieve these inversions. Furthermore, several atomic configurations had to be tested for each inversion to find the exact atomic configuration of the system. In particular, for inversions corresponding to x = 0.67, 0.75, and 1, we carried out 54, 38, and 2 calculations, respectively. Therefore, 94 single-point calculations were carried out to determine the most stable atomic configurations. During these calculations, we did not add any restrictions to the spin-polarized solutions. Once the most stable atomic configurations were determined, we ran a final geometry optimization. The final bulk geometries are shown in **Figure S13**, and crystallographic data is found at the end of this supplementary document.

**Slab calculations**

From **Fig. S9** it was determined that ZFO NPs are primarily octahedral in morphology, indicating the predominance of (111) surface planes. On the other hand, CFO and NFO NPs most closely resemble truncated octahedra, which would indicate the presence of dominating (111) and (100) planes. Slabs were built by cutting the previously determined bulk spinels along these crystallographic directions. This was achieved with the keyword SLABCUT within the CRYSTAL suite. By default, 2D slabs in CRYSTAL17 are built with a distance of 500 Å between periodic images of the structure in the direction perpendicular to the SLABCUT planes. This avoids unwanted interactions between slabs in the z-direction. Compared to the use of plane waves, Gaussian orbitals allow for this large vacuum space since they are localized and do not increase the computational demand upon the addition of vacuum. Finally, hydrogen and oxygen atoms were added to "cap" both sides of the slabs due to incomplete bonding, exposure of transition metals at the surface, and charge imbalances. The final structures are shown in **Fig. 3** of the main manuscript and **Fig. S14** of the S.I. Furthermore, crystallographic data is shown at the end of this supplementary document.

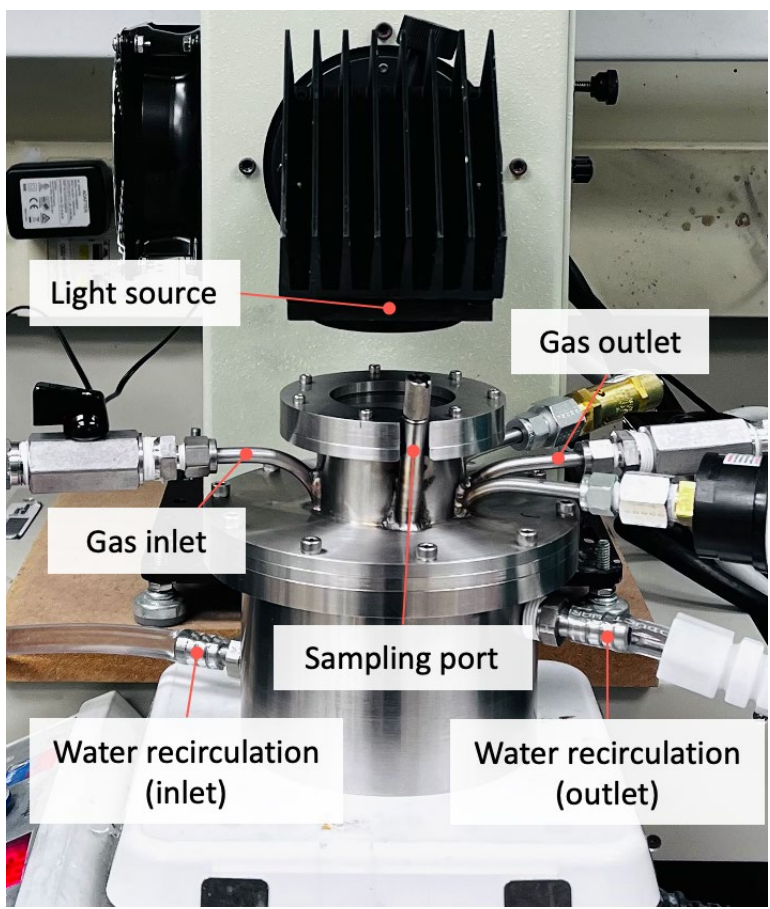

**Figure S1.** Custom-built photoreactor with a temperature control system (water recirculation).

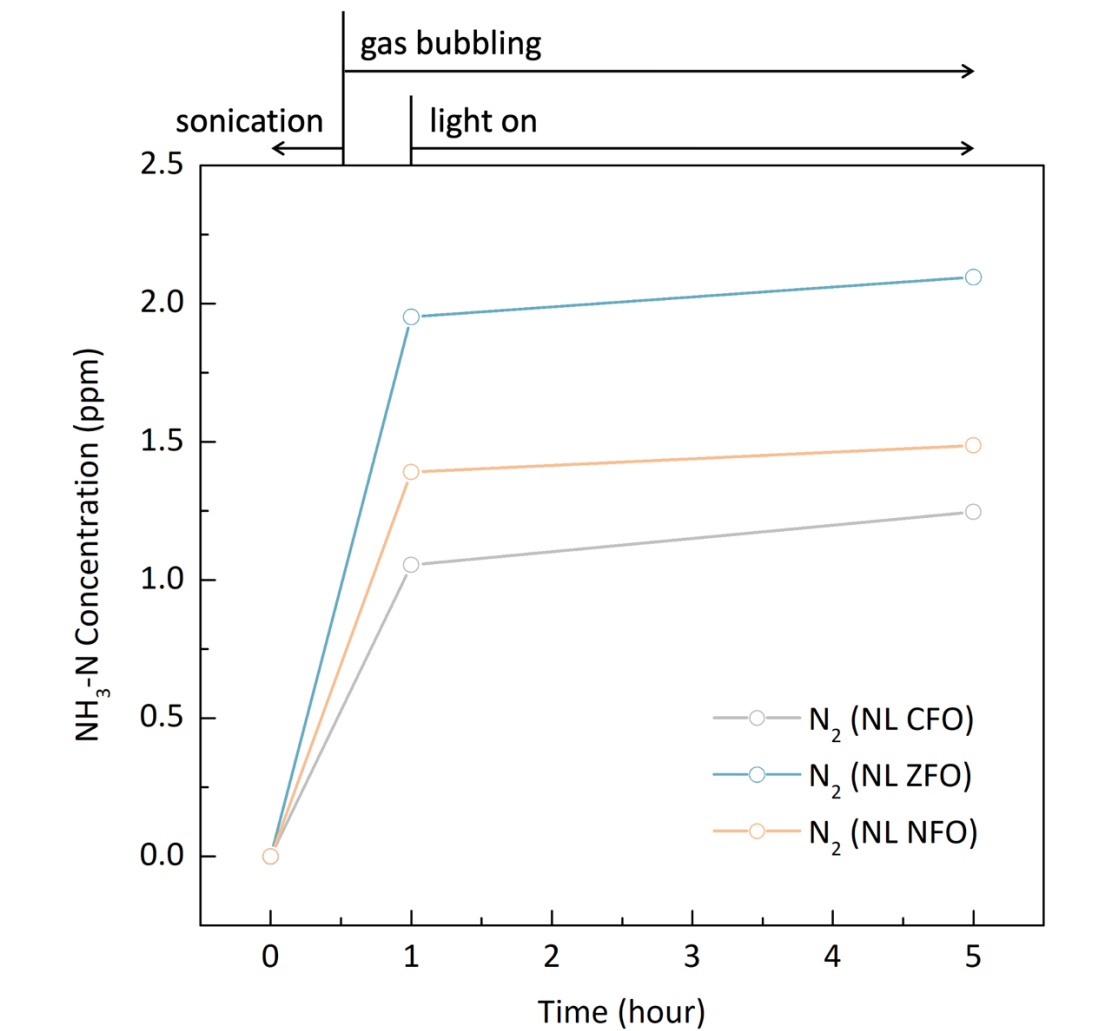

**Figure S2.** Time course of NH$_3$ evolution for NL Co, Zn, and Ni ferrite under full-spectrum irradiation in N$_2$ environment.

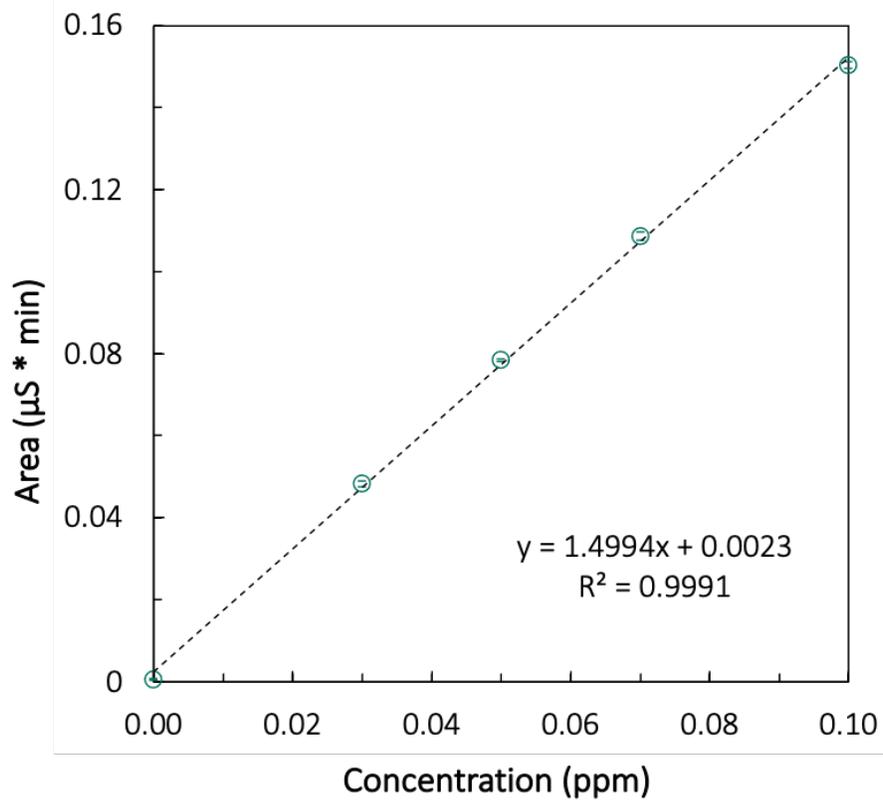

**Figure S3.** Calibration curves for ammonia measurement using ion chromatography. The error bars represent the standard deviation from three independent measurements.

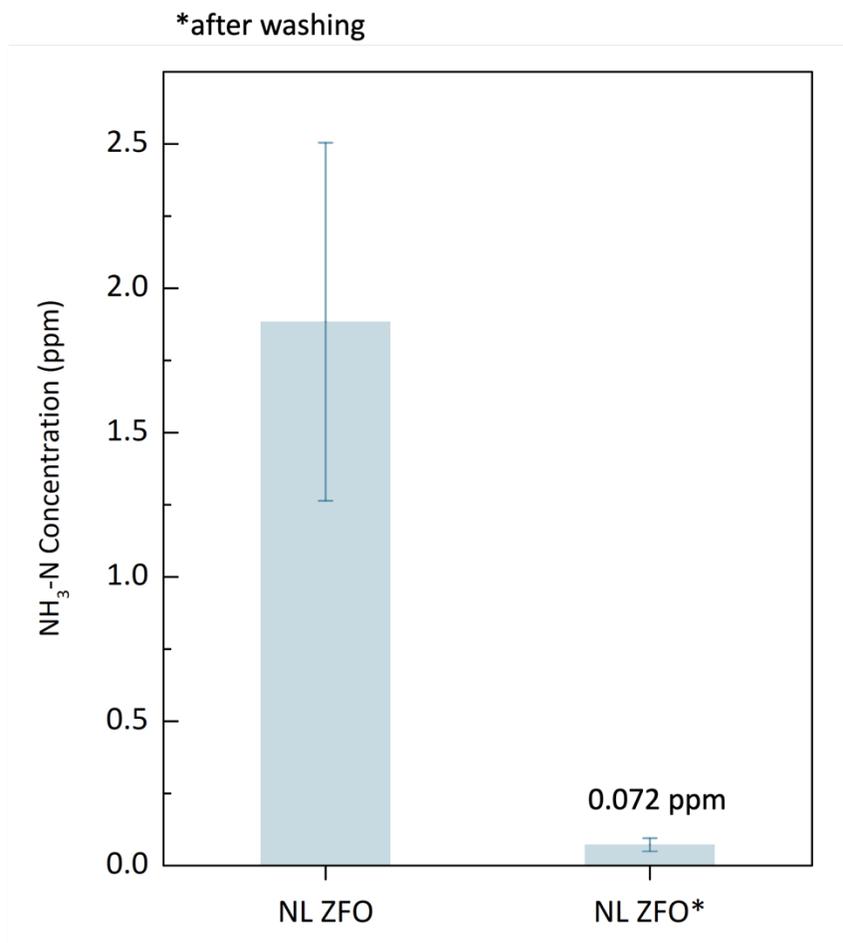

**Figure S4.** A comparison of ammonia in solutions of NL ZnFeO4 (ZFO) before and after washing. The sample was collected after 30 minutes of sonication (catalyst in DI water) in the absence of illumination. The error bars represent the standard deviation from three independent experiments.

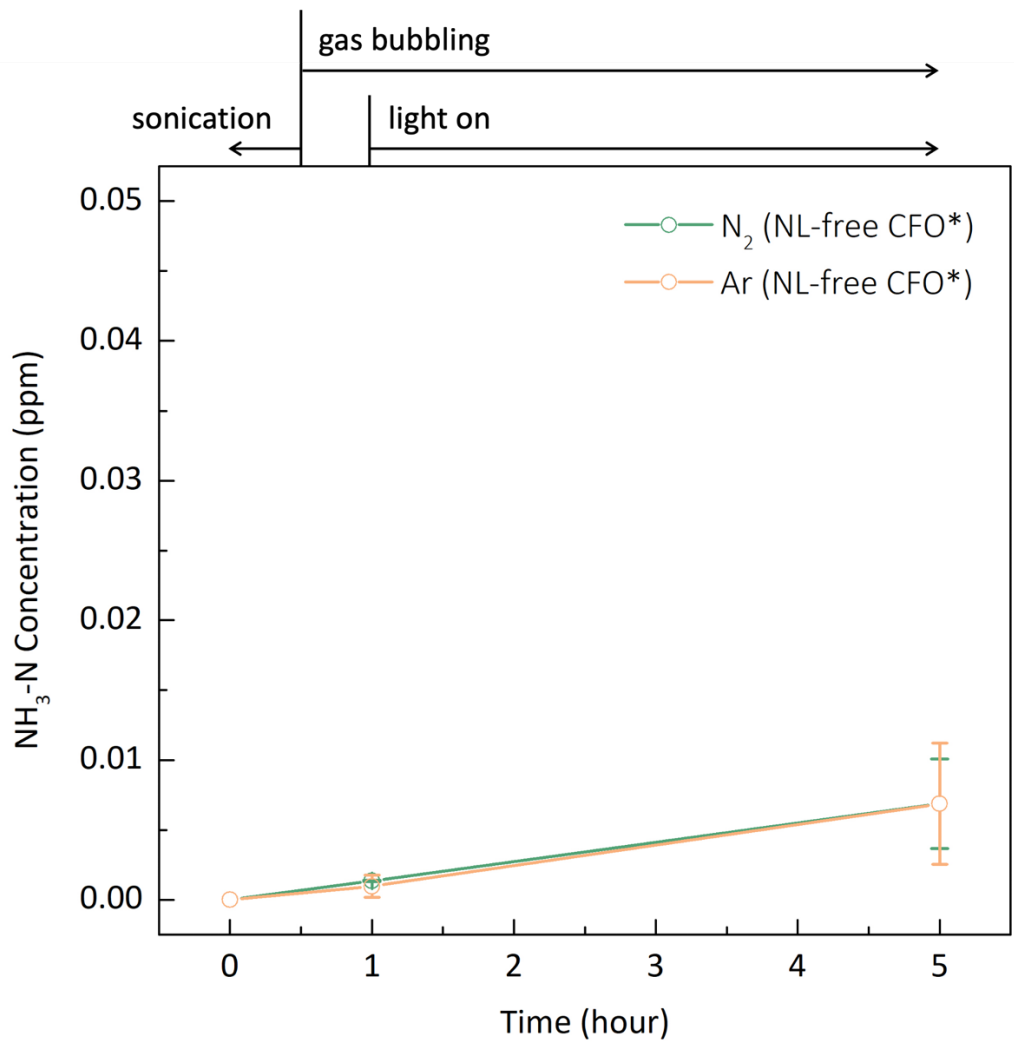

**Figure S5.** Time course of NH₃ evolution for NL-free CoFeO₄ (CFO) under full-spectrum irradiation in N₂ or Ar environments (raw data of **Fig. 2**). The photocatalytic activity reported in **Fig. 2** was obtained by subtracting the first-hour data point from the fifth-hour data point. The error bars represent the standard deviation from two independent experiments.

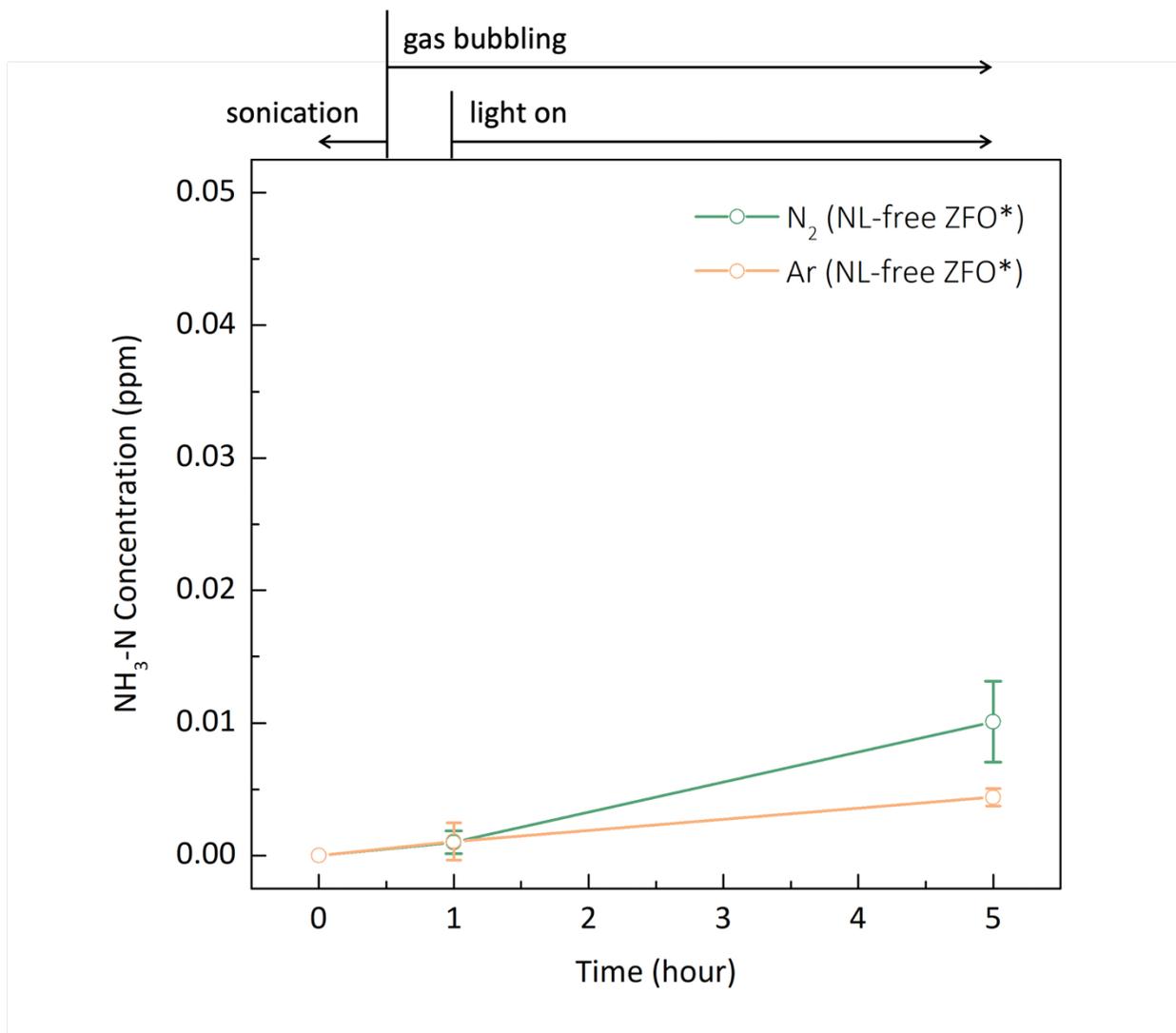

**Figure S6.** Time course of NH$_3$ evolution for NL-free ZnFeO$_4$ (ZFO) under full-spectrum irradiation in N$_2$ or Ar environments (raw data of **Fig. 2**). The photocatalytic activity reported in **Fig. 2** was obtained by subtracting the first-hour data point from the fifth-hour data point. The error bars represent the standard deviation from two independent experiments.

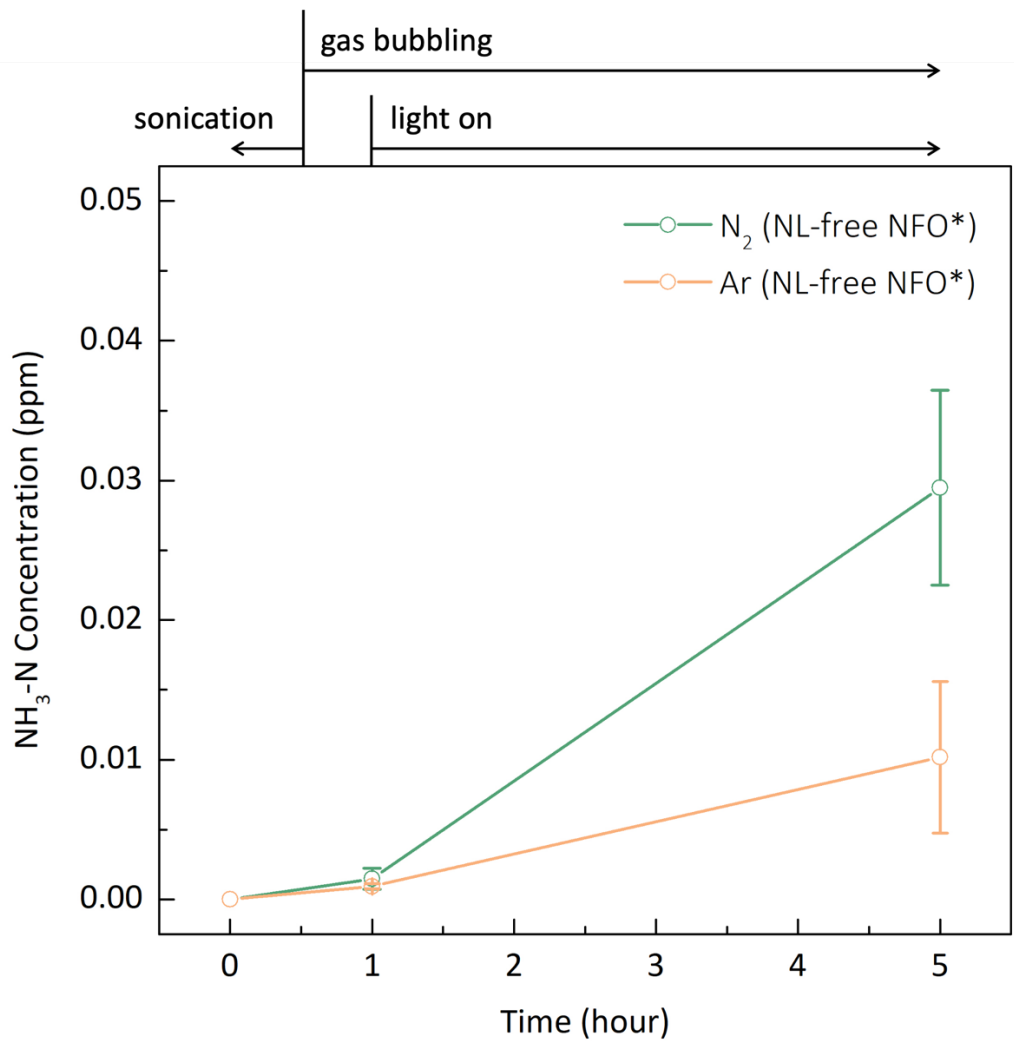

**Figure S7.** Time course of NH$_3$ evolution for NL-free NiFeO$_4$ (NFO) under full-spectrum irradiation in N$_2$ or Ar environments (raw data of **Fig. 2**). The photocatalytic activity reported in **Fig. 2** was obtained by subtracting the first-hour data point from the fifth-hour data point. The error bars represent the standard deviation from two independent experiments.

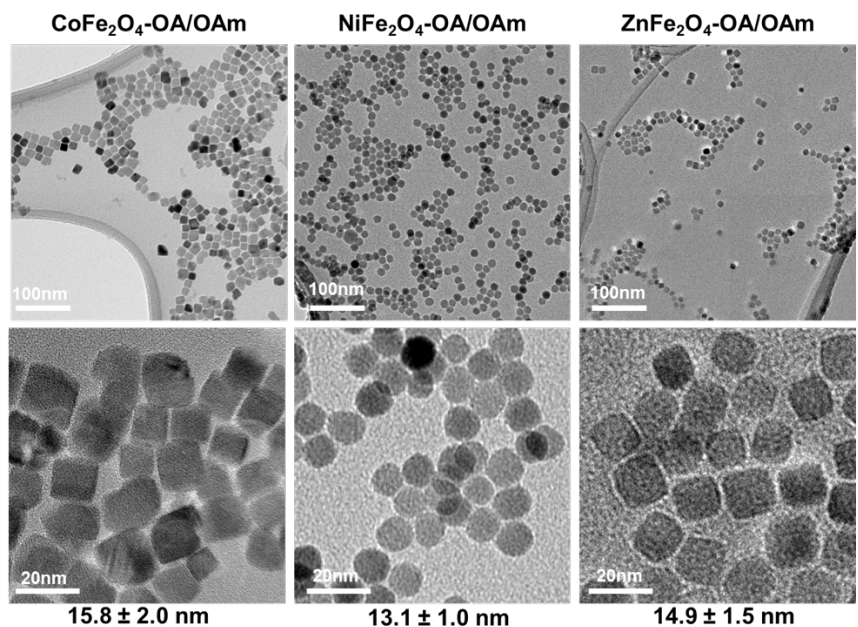

**Figure S8.** Representative TEM micrographs of NL Co, Ni and Zn ferrites with OA and OAm as surface ligands.

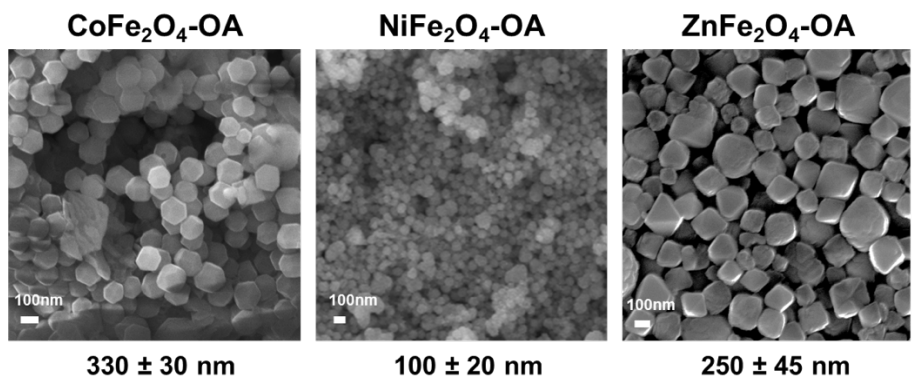

**Figure S9.** Representative SEM micrographs of NL-free Co, Ni and Zn ferrites with OA as surface ligand.

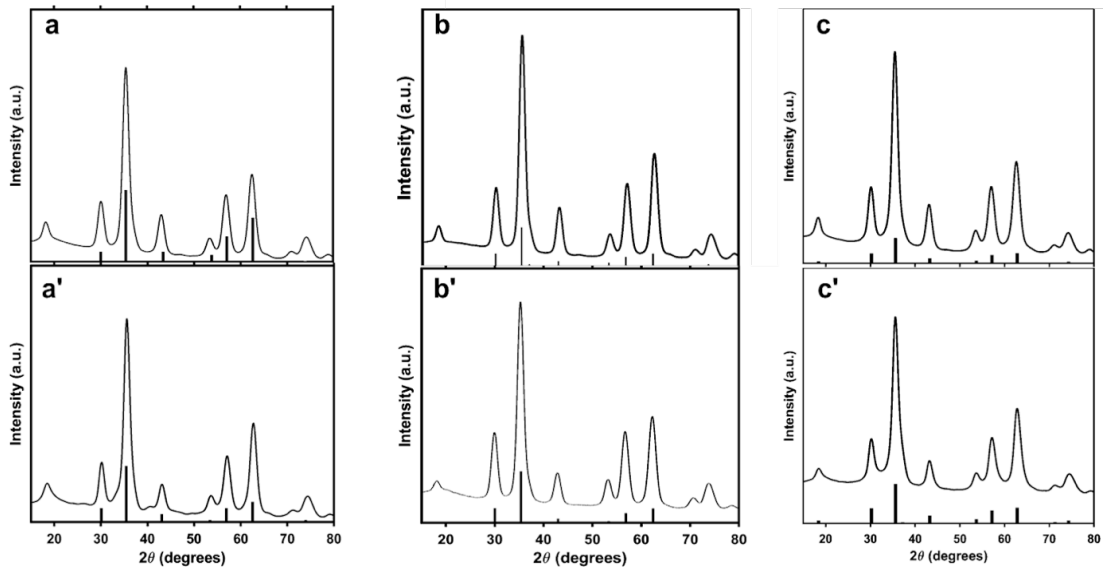

**Figure S10.** Representative XRD diffractograms of (a, b and c) NL Co, Ni, Zn ferrites with OA/OAm, and (a', b', c') NL-free Co, Ni and Zn ferrites with OA as surface ligands. JCPDS: 03-0864 ($CoFe_2O_4$), 82-1049 ($ZnFe_2O_4$) and 10-0325 ($NiFe_2O_4$).

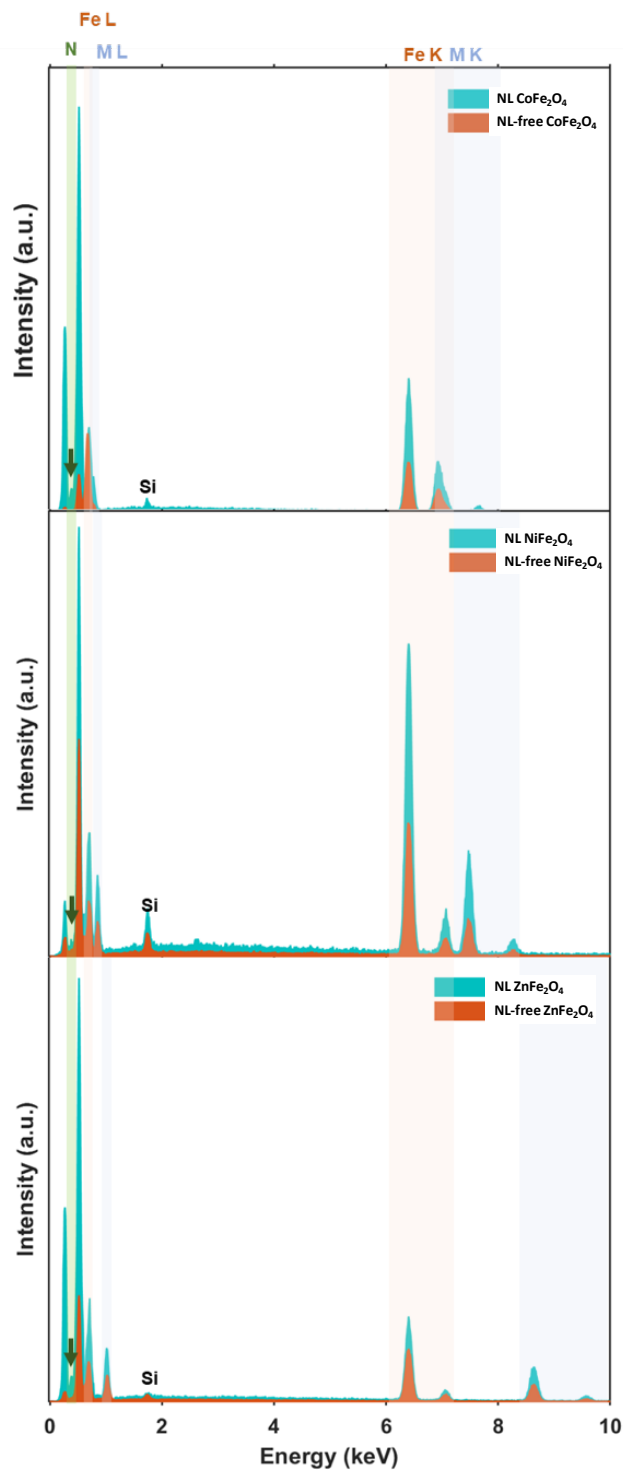

**Figure S11.** Representative EDS spectra from NL and NL-free ferrites showcasing the potential presence of the nitrogen K-alpha signal at 0.39 keV. The position of the N peak (N K-alpha) lies between the C and the O K-alpha. It is very difficult to discern the peak due to limited resolution: The total concentration of N should be much lower than that of C and O and N possesses a low Z number.

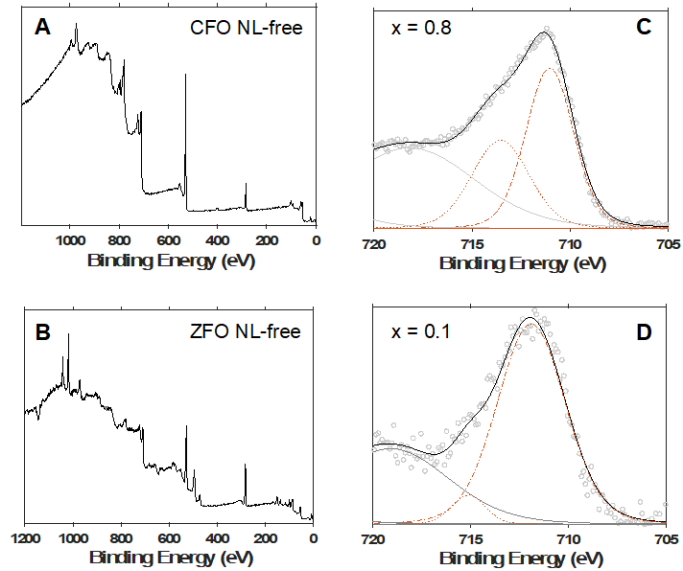

**Figure S12. A,B)** Survey XPS scans of nitrogen ligand-free CFO (**A**) and ZFO (**B**) nanocrystals indicating the presence of C, O, Fe, and Co, or Zn, respectively. **C,D)** Detail XPS scans of the Fe $2p_{3/2}$ peaks for nitrogen ligand-free CFO (C) and ZFO (D) nanocrystals. These peaks each contain two components highlighted in orange: the component with higher binding energy corresponds to $Fe^{3+}$ in tetrahedral sites and the component with lower binding energy correspond to $Fe^{3+}$ in octahedral sites. The inversion parameters listed in the insets are determined from the areas of these two peaks using the equation $x = 2\frac{A_{Fe_{tet}}}{A_{Fe_{tet}}+A_{Fe_{oct}}}$.[7]

**2.**

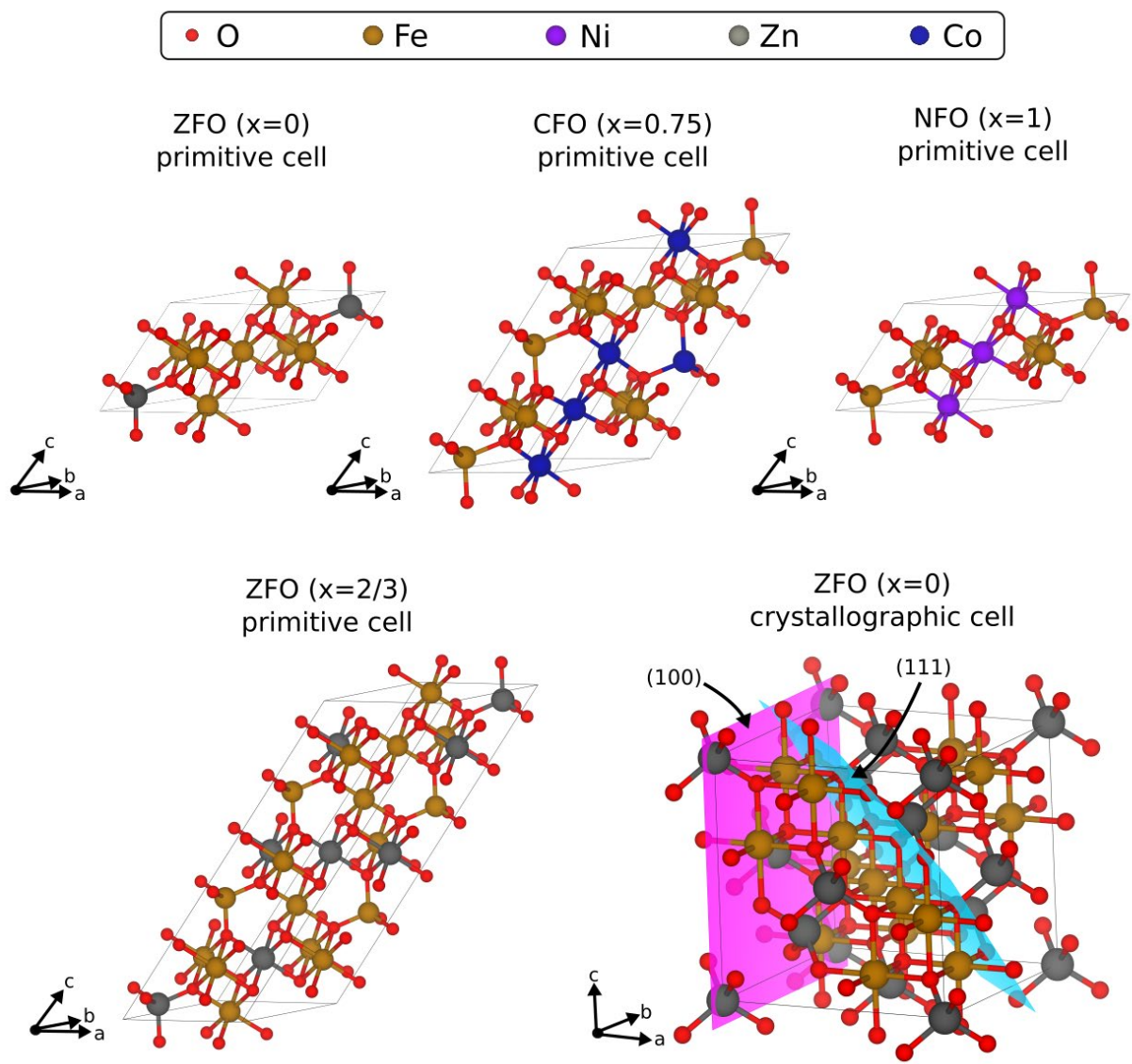

**Figure S13.** ZFO, NFO, and CFO bulk primitive cells used in this work. Note that the CFO 0.75 inversion required a 1✕1✕2 supercell and the ZFO 2/3 inversion required a 1✕1✕3 supercell. The bottom right subfigure shows an example of a normal ZFO spinel crystallographic cell. The (100) and (111) planes are drawn to show the planes normal to the direction in which slabs were cut.

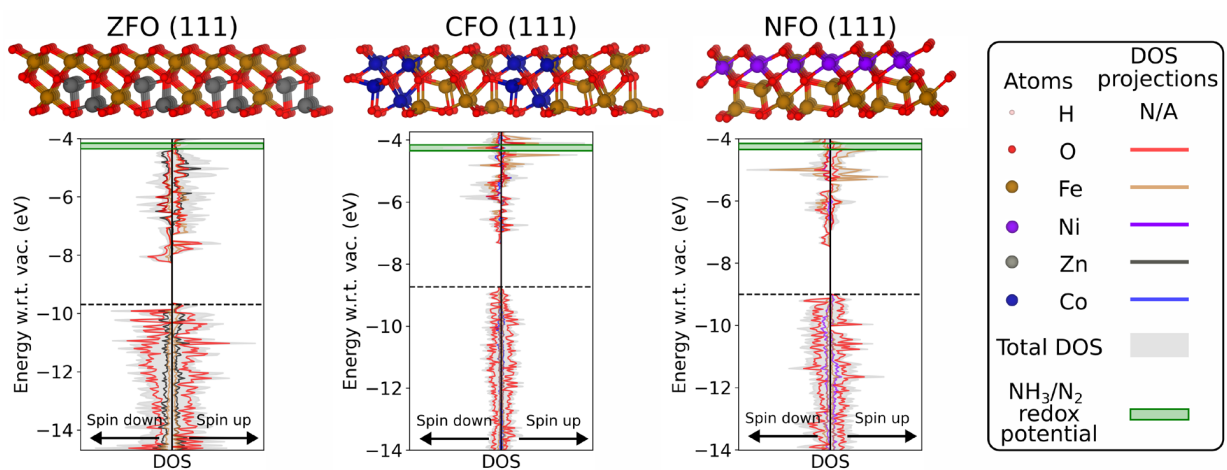

**Figure S14.** Spin-polarized, atom projected electronic density of states for ZFO, CFO, and NFO (111) slabs without surface hydrogen. These plots are aligned with respect to vacuum, and the $NH_3/N_2$ potential is plotted as a green rectangle. It can be observed that the structures have their CBM below the energy range of interest, indicating that ammonia generation is not feasible.

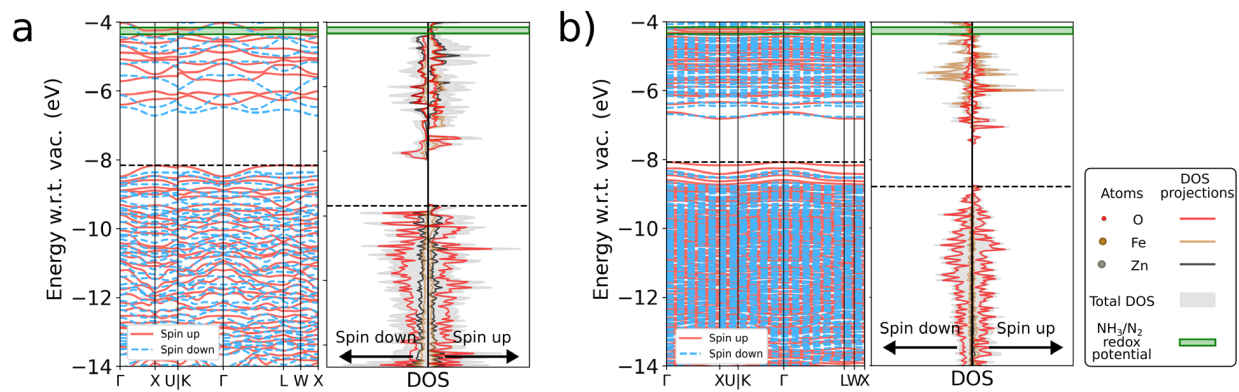

**Figure S15** Spin-polarized electronic properties (band structures and density of states) for **a)** normal ($x=0$) and **b)** inverted ($x=2/3$) ZFO (111) slabs. The electronic properties are aligned with respect to vacuum, and the $NH_3/N_2$ potential is plotted as a green rectangle. It can be observed that neither structure has their CBM above the energy range of interest, indicating that ammonia generation is not feasible.

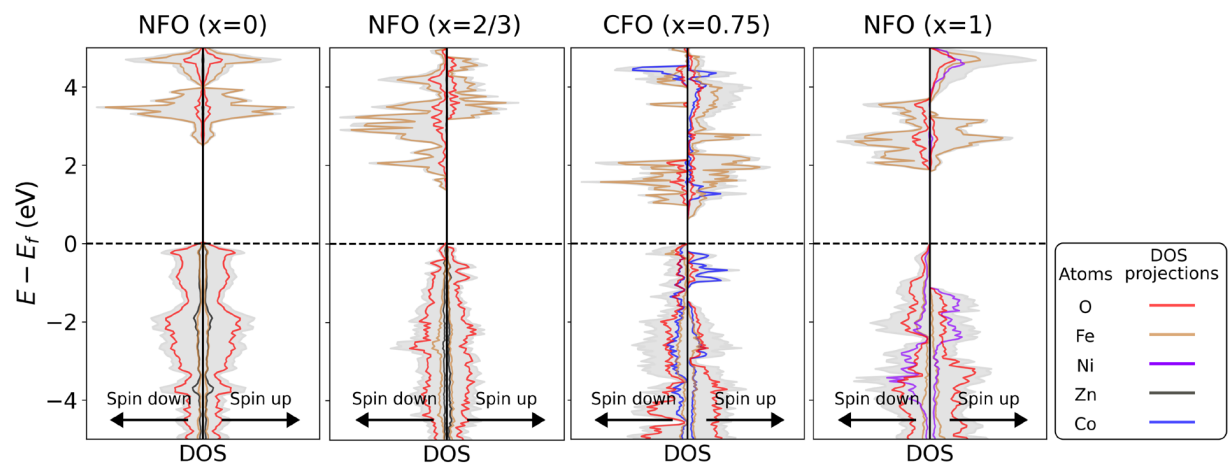

**Figure S16** Spin-polarized, atom projected electronic density of states for ZFO, CFO, and NFO bulk structures.

*Crytallographic Structures in CIF format*

## ZFO bulk normal spinel

```
data_ZnFe2O4_bulk_normal

_cell_length_a                          5.929088
_cell_length_b                          5.945991
_cell_length_c                          5.945486
_cell_angle_alpha                       59.817862
_cell_angle_beta                        60.098437
_cell_angle_gamma                       60.094733
_symmetry_space_group_name_H-M          'P 1'
_symmetry_Int_Tables_number             1

loop_
_symmetry_equiv_pos_as_xyz
   'x, y, z'

loop_
   _atom_site_label
   _atom_site_type_symbol
   _atom_site_fract_x
   _atom_site_fract_y
   _atom_site_fract_z
Zn001   Zn    0.124972    0.125010    0.125008
Zn002   Zn   -0.125006   -0.124973   -0.125009
Fe003   Fe   -0.499841   -0.499991    0.499842
Fe004   Fe    0.499950    0.499880    0.000018
Fe005   Fe    0.000100   -0.499972    0.499989
Fe006   Fe    0.499945   -0.000054   -0.499865
 O007    O    0.259252    0.261023    0.261064
 O008    O    0.261070    0.259343   -0.281352
 O009    O   -0.281377    0.261001    0.261033
 O010    O    0.261015   -0.281281    0.259258
 O011    O   -0.259329   -0.261083   -0.260918
 O012    O   -0.261011   -0.259225    0.281353
 O013    O   -0.261007    0.281360   -0.259354
 O014    O    0.281267   -0.261039   -0.261067
```

## ZFO bulk 2/3 inversion spinel

```
data_ZnFe2O4_bulk_inv

_cell_length_a                          5.843106
_cell_length_b                          5.843106
_cell_length_c                          17.529318
_cell_angle_alpha                       60.000000
_cell_angle_beta                        60.000000
_cell_angle_gamma                       60.000000
_symmetry_space_group_name_H-M          'P 1'
_symmetry_Int_Tables_number             1

loop_
_symmetry_equiv_pos_as_xyz
   'x, y, z'
```

```
loop_
   _atom_site_label
   _atom_site_type_symbol
   _atom_site_fract_x
   _atom_site_fract_y
   _atom_site_fract_z
Zn001   Zn    0.125000    0.125000    0.041667
Fe002   Fe    0.125000    0.125000    0.375000
Fe003   Fe    0.125000    0.125000   -0.291667
Zn004   Zn   -0.125000   -0.125000   -0.041667
Fe005   Fe   -0.125000   -0.125000    0.291667
Fe006   Fe   -0.125000   -0.125000   -0.375000
Zn007   Zn    0.500000   -0.500000    0.166667
Zn008   Zn    0.500000    0.500000   -0.500000
Fe009   Fe   -0.500000   -0.500000   -0.166667
Fe010   Fe   -0.500000   -0.500000    0.000000
Fe011   Fe   -0.500000    0.500000    0.333333
Fe012   Fe   -0.500000   -0.500000   -0.333333
Fe013   Fe   -0.000000    0.500000    0.166667
Zn014   Zn   -0.000000   -0.500000   -0.500000
Zn015   Zn   -0.000000   -0.500000   -0.166667
Fe016   Fe    0.500000    0.000000    0.166667
Fe017   Fe    0.500000   -0.000000   -0.500000
Fe018   Fe   -0.500000   -0.000000   -0.166667
O019    O     0.249000    0.249000    0.083000
O020    O     0.249000    0.249000    0.416333
O021    O     0.249000    0.249000   -0.250333
O022    O     0.249000    0.249000   -0.082333
O023    O     0.249000    0.249000    0.251000
O024    O     0.249000    0.249000   -0.415667
O025    O    -0.247000    0.249000    0.083000
O026    O    -0.247000    0.249000    0.416333
O027    O    -0.247000    0.249000   -0.250333
O028    O     0.249000   -0.247000    0.083000
O029    O     0.249000   -0.247000    0.416333
O030    O     0.249000   -0.247000   -0.250333
O031    O    -0.249000   -0.249000   -0.083000
O032    O    -0.249000   -0.249000    0.250333
O033    O    -0.249000   -0.249000   -0.416333
O034    O    -0.249000   -0.249000    0.082333
O035    O    -0.249000   -0.249000    0.415667
O036    O    -0.249000   -0.249000   -0.251000
O037    O    -0.249000    0.247000   -0.083000
O038    O    -0.249000    0.247000    0.250333
O039    O    -0.249000    0.247000   -0.416333
O040    O     0.247000   -0.249000   -0.083000
O041    O     0.247000   -0.249000    0.250333
O042    O     0.247000   -0.249000   -0.416333
```

### CFO bulk 0.75 inversion spinel

```
data_CoFe2O4_bulk_inv

_cell_length_a                          5.732085
_cell_length_b                          5.809447
```

```
_cell_length_c                              11.550516
_cell_angle_alpha                           59.396031
_cell_angle_beta                            60.625375
_cell_angle_gamma                           60.268693
_symmetry_space_group_name_H-M              'P 1'
_symmetry_Int_Tables_number                 1

loop_
_symmetry_equiv_pos_as_xyz
   'x, y, z'

loop_
   _atom_site_label
   _atom_site_type_symbol
   _atom_site_fract_x
   _atom_site_fract_y
   _atom_site_fract_z
Fe001  Fe   0.125330   0.123437   0.060785
Fe002  Fe   0.125330   0.123437  -0.439215
Co003  Co  -0.125354  -0.123433  -0.060769
Fe004  Fe  -0.125354  -0.123433   0.439231
Fe005  Fe  -0.499972   0.499993  -0.249980
Fe006  Fe  -0.499972   0.499993   0.250020
Co007  Co   0.499999   0.499983  -0.000019
Fe008  Fe   0.499999   0.499983   0.499981
Fe009  Fe   0.000012  -0.499928   0.249967
Co010  Co   0.000012  -0.499928  -0.250033
Fe011  Fe  -0.499977  -0.000000   0.249999
Co012  Co  -0.499977  -0.000000  -0.250001
 O013   O   0.274984   0.253941   0.125699
 O014   O   0.274984   0.253941  -0.374301
 O015   O   0.267167   0.261939  -0.147111
 O016   O   0.267167   0.261939   0.352889
 O017   O  -0.289769   0.262051   0.134544
 O018   O  -0.289769   0.262051  -0.365456
 O019   O   0.265471  -0.291180   0.129019
 O020   O   0.265471  -0.291180  -0.370981
 O021   O  -0.274925  -0.253931  -0.125708
 O022   O  -0.274925  -0.253931   0.374292
 O023   O  -0.267078  -0.261936   0.147077
 O024   O  -0.267078  -0.261936  -0.352923
 O025   O  -0.265564   0.291060  -0.128965
 O026   O  -0.265564   0.291060   0.371035
 O027   O   0.289675  -0.261996  -0.134536
 O028   O   0.289675  -0.261996   0.365464
```

## NFO bulk 1.0 inversion spinel

```
data_NiFe2O4_bulk_inv

_cell_length_a 5.84081513
_cell_length_b 5.82785058
_cell_length_c 5.68620063
_cell_angle_alpha 60.459571
_cell_angle_beta 60.249354
_cell_angle_gamma 57.520708
```

```
_symmetry_space_group_name_H-M          'P 1'
_symmetry_Int_Tables_number             1

loop_
_symmetry_equiv_pos_as_xyz
   'x, y, z'

loop_
_atom_site_label
_atom_site_type_symbol
_atom_site_fract_x
_atom_site_fract_y
_atom_site_fract_z
Fe001 Fe  1.191755771790E-01  1.189306951760E-01  1.259628145440E-01
Fe002 Fe -1.176447274740E-01 -1.183967929360E-01 -1.270155504000E-01
Ni003 Ni  4.997847882590E-01  4.998000348240E-01  4.995354131640E-01
Ni004 Ni  4.983670684070E-01  4.983110732740E-01  4.364895859000E-03
Fe005 Fe  3.377899611000E-03  4.986900031680E-01  4.984765806570E-01
Fe006 Fe  4.987951837740E-01  3.615800683000E-03  4.983240208230E-01
 O007  O   2.583015731800E-01  2.563242431490E-01  2.626084172350E-01
 O008  O   2.598308720040E-01  2.624786511890E-01 -2.867646270770E-01
 O009  O  -2.906079079160E-01  2.413823227950E-01  2.584409392350E-01
 O010  O   2.414635550840E-01 -2.909860474440E-01  2.602171183220E-01
 O011  O  -2.546418666490E-01 -2.522388735890E-01 -2.597946211610E-01
 O012  O  -2.540071179450E-01 -2.558903050460E-01  2.705742876240E-01
 O013  O  -2.410195050620E-01  2.786296172560E-01 -2.531625492840E-01
 O014  O   2.788246075470E-01 -2.406504224980E-01 -2.517671395400E-01
```

*Note that **slab** CIF files have a c lattice parameter of 40 or 60 Å for better visualization in the VESTA software. However, calculations were carried out with a cell length of 500 Å in the c-direction to avoid interactions between periodic images of the slab. This spacing is automatically assigned by the CRYSTAL17 software when setting up slab calculations.*

### (111) ZFO slab 2/3 inversion spinel (without H)

```
data_ZnFe2O4_111_slab_inv_noH

_cell_length_a                         5.956608
_cell_length_b                         15.534336
_cell_length_c                         40.000000
_cell_angle_alpha                      90.000000
_cell_angle_beta                       90.000000
_cell_angle_gamma                      101.471900
_symmetry_space_group_name_H-M         'P 1'
_symmetry_Int_Tables_number            1

loop_
_symmetry_equiv_pos_as_xyz
   'x, y, z'

loop_
   _atom_site_label
   _atom_site_type_symbol
   _atom_site_fract_x
   _atom_site_fract_y
   _atom_site_fract_z
 O001    O    0.177282   0.129196   0.049827
 O002    O   -0.446709  -0.190130   0.057455
 O003    O   -0.130593   0.462848   0.054898
 O004    O   -0.148453  -0.044073   0.057076
 O005    O    0.226608  -0.382208   0.055604
 O006    O   -0.465017   0.288861   0.055206
 O007    O   -0.310995   0.131360   0.048598
 O008    O   -0.002577  -0.192640   0.057457
 O009    O    0.345320   0.469951   0.054072
 O010    O    0.303501  -0.018700   0.058002
 O011    O   -0.325690  -0.364874   0.061250
 O012    O   -0.023222   0.277720   0.051693
 Zn013   Zn   0.308930  -0.135239   0.035522
 Zn014   Zn  -0.345678  -0.472524   0.034163
 Fe015   Fe  -0.042384   0.189717   0.025956
 Fe016   Fe   0.119583   0.027366   0.027100
 Fe017   Fe   0.474426  -0.304221   0.035556
 Fe018   Fe  -0.198371   0.357411   0.031240
 Fe019   Fe  -0.202384  -0.136303   0.034630
 Fe020   Fe   0.134186  -0.477210   0.032029
 Fe021   Fe   0.453747   0.186763   0.025077
 O022    O    0.381353   0.079220  -0.003656
 O023    O   -0.254252  -0.245852   0.005764
```

```
O024   O    0.088342   0.418164   0.002441
O025   O   -0.405882  -0.097231   0.004331
O026   O   -0.054405  -0.421341   0.002360
O027   O    0.217822   0.240566  -0.002117
O028   O   -0.077146   0.080880   0.000206
O029   O    0.274955  -0.246537   0.004088
O030   O   -0.401384   0.417545   0.001076
O031   O    0.055008  -0.084097   0.006016
O032   O    0.438759  -0.413067  -0.000640
O033   O   -0.278111   0.247715  -0.003583
Zn034  Zn  -0.380460   0.017120  -0.018173
Fe035  Fe  -0.012970  -0.301932  -0.009365
Fe036  Fe   0.318658   0.361849  -0.013057
Fe037  Fe  -0.461584  -0.196708  -0.029323
Zn038  Zn  -0.128421   0.479727  -0.033055
Zn039  Zn   0.169138   0.128836  -0.039346
Zn040  Zn   0.081715  -0.093900  -0.044334
Fe041  Fe   0.436092  -0.412156  -0.043491
Fe042  Fe  -0.326053   0.265582  -0.049162
O043   O   -0.301442   0.093829  -0.062041
O044   O    0.002213  -0.250979  -0.048546
O045   O    0.419582   0.341384  -0.051947
O046   O    0.418655  -0.094275  -0.046909
O047   O   -0.340757  -0.448542  -0.055506
O048   O   -0.136997   0.168301  -0.059214
O049   O    0.104965   0.019763  -0.065804
O050   O    0.451650  -0.309119  -0.053548
O051   O   -0.159234   0.375240  -0.063259
O052   O   -0.211186  -0.157605  -0.060066
O053   O    0.196137  -0.473034  -0.054889
O054   O    0.413728   0.195582  -0.068395
```

**(111) ZFO slab normal spinel (without H)**

**data_ZnFe2O4_111_slab_normal_noH**

**_cell_length_a 5.67104688**
**_cell_length_b 5.70012773**
**_cell_length_c 40**
**_cell_angle_alpha 90.000000**
**_cell_angle_beta 90.000000**
**_cell_angle_gamma 119.904400**
**_symmetry_space_group_name_H-M          'P 1'**
**_symmetry_Int_Tables_number             1**

**loop_**
**_symmetry_equiv_pos_as_xyz**
   **'x, y, z'**

**loop_**
**_atom_site_label**
**_atom_site_type_symbol**
**_atom_site_fract_x**
**_atom_site_fract_y**
**_atom_site_fract_z**
 O001   O    4.119276665586E-01   3.294715990367E-01   0.053494415944375

```
 O002  O -1.014368909290E-01 -1.677690069143E-01  0.051942704758475
 O003  O  4.295856194160E-01 -1.559144978354E-01  0.052419416458325
 O004  O -9.529368609301E-02  3.042906342187E-01  0.052705572849500004
Fe005 Fe -4.258815720366E-01  1.510384777069E-01  0.03153296195815
Fe006 Fe  7.369255655706E-02  1.679263048898E-01  0.0312438590094
Fe007 Fe -4.098958037754E-01 -3.277146779273E-01  0.03181061822185
 O008  O  2.676389401613E-01  4.723014238107E-01  0.0015492454626895
 O009  O -2.938597484588E-01  4.827637909968E-01  0.00082581113721025
 O010  O  2.648623006905E-01  4.155941077182E-02  6.994066739035e-05
 O011  O -2.558490805426E-01 -3.989694868562E-03  0.0048543842903449996
Zn012 Zn  8.442207467449E-02 -3.372478977164E-01 -0.0106560955060875
Fe013 Fe  4.266245006948E-01  3.387933098556E-01 -0.031351390927174996
Zn014 Zn -2.690651078714E-01 -1.009109103158E-02 -0.044038239292325
 O015  O  1.638721099410E-01 -3.047993677177E-01 -0.06099215049534999
 O016  O -3.914793777127E-01  2.382002996182E-01 -0.05465052409397499
 O017  O  1.100757534867E-01  1.360025652211E-01 -0.0575798955803
 O018  O  4.980597452389E-01 -3.788215821152E-01 -0.056758791915775
```

## (111) ZFO slab normal spinel with H on top [octahedral] layer

**data_ZnFe2O4_111_slab_normal_Htop**

| | |
|---|---|
| _cell_length_a | 5.91426397 |
| _cell_length_b | 5.99201685 |
| _cell_length_c | 60.00000000 |
| _cell_angle_alpha | 90.000000 |
| _cell_angle_beta | 90.000000 |
| _cell_angle_gamma | 119.497300 |
| _symmetry_space_group_name_H-M | 'P 1' |
| _symmetry_Int_Tables_number | 1 |

```
loop_
_symmetry_equiv_pos_as_xyz
   'x, y, z'

loop_
   _atom_site_label
   _atom_site_type_symbol
   _atom_site_fract_x
   _atom_site_fract_y
   _atom_site_fract_z
 O001   O   0.4100292465   0.3359475486   0.0392948707
 O002   O  -0.1021225316  -0.1509034140   0.0398260076
 O003   O   0.4171738958  -0.1562590583   0.0396294731
 O004   O  -0.1026626871   0.3212555402   0.0396064184
Fe005  Fe  -0.4247684671   0.1619918301   0.0199701331
Fe006  Fe   0.0697655221   0.1694723552   0.0199502029
Fe007  Fe  -0.4246070498  -0.3268310262   0.0203910740
 O008   O   0.2540967609   0.4943234905   0.0033815457
 O009   O  -0.2993688087   0.4758904351   0.0017189595
 O010   O   0.2601013099   0.0386965403   0.0019733908
 O011   O  -0.2562264182  -0.0014672214   0.0034971651
Zn012  Zn   0.0752270919  -0.3304574290  -0.0084180055
Fe013  Fe   0.4179017703   0.3354971901  -0.0214207227
Zn014  Zn  -0.2272869819  -0.0145007477  -0.0310820484
 O015   O  -0.0108918679  -0.2725939000  -0.0385717643
```

```
O016   O   -0.4108847174   0.1699742773   -0.0374448490
O017   O    0.1405571769   0.2186439016   -0.0357598464
O018   O   -0.4227739108  -0.3710384115   -0.0375731719
H019   H    0.4137114761   0.3340997046    0.0553528410
H020   H   -0.0017816915  -0.2403215580    0.0409708569
H021   H   -0.4965976921  -0.1265185382    0.0539046771
H022   H   -0.1265914264   0.2390984906    0.0539541940
```

## (111) ZFO slab normal spinel with H on bottom [octahedral+tetrahedral] layer

```
data_ZnFe2O4_111_slab_normal_Hbot

_cell_length_a                            5.97540120
_cell_length_b                            5.84634139
_cell_length_c                            60.00000000
_cell_angle_alpha                         90.000000
_cell_angle_beta                          90.000000
_cell_angle_gamma                         121.012500
_symmetry_space_group_name_H-M            'P 1'
_symmetry_Int_Tables_number               1

loop_
_symmetry_equiv_pos_as_xyz
   'x, y, z'

loop_
   _atom_site_label
   _atom_site_type_symbol
   _atom_site_fract_x
   _atom_site_fract_y
   _atom_site_fract_z
 O001   O    0.4100236134   0.3452518672    0.0347185659
 O002   O   -0.0696515041  -0.1165827469    0.0360755358
 O003   O    0.4367448626  -0.1741595865    0.0326290318
 O004   O   -0.1294303541   0.2865805069    0.0349697573
Fe005   Fe  -0.4391883157   0.1350382864    0.0193016050
Fe006   Fe   0.0808411994   0.1522609384    0.0192029286
Fe007   Fe  -0.4109045382  -0.3445186244    0.0193221776
 O008   O    0.2459790704   0.4368939077   -0.0011110948
 O009   O   -0.2846942720   0.4692468945    0.0018927762
 O010   O    0.2844706351   0.0126069311   -0.0003825912
 O011   O   -0.2385256330  -0.0390159540    0.0007602974
Zn012   Zn   0.0796923216  -0.3635834241   -0.0092700754
Fe013   Fe   0.4155371649   0.3082538323   -0.0246822514
Zn014   Zn  -0.2544223878  -0.0289013834   -0.0324396870
 O015   O    0.0826527746  -0.3717887577   -0.0415628519
 O016   O   -0.3636378120   0.2164659394   -0.0402993827
 O017   O    0.0972533768   0.0945917565   -0.0432579929
 O018   O   -0.4985548724  -0.3832430659   -0.0421757465
H019    H    0.0989185132  -0.2056325558   -0.0470048129
H020    H   -0.1925826519   0.3898379937   -0.0438964082
H021    H    0.1345504771   0.1949895672   -0.0570159030
H022    H    0.3335949987  -0.3772589892   -0.0440142140
```

## (111) CFO slab 0.75 inversion spinel (without H)

```
data_CoFe2O4_111_slab_inv_noH

_cell_length_a 5.83022889
_cell_length_b 9.94655311
_cell_length_c 40
_cell_angle_alpha 90.000000
_cell_angle_beta 90.000000
_cell_angle_gamma 92.101100
_symmetry_space_group_name_H-M          'P 1'
_symmetry_Int_Tables_number             1

loop_
_symmetry_equiv_pos_as_xyz
   'x, y, z'

loop_
_atom_site_label
_atom_site_type_symbol
_atom_site_fract_x
_atom_site_fract_y
_atom_site_fract_z
 O001   O  -3.948743380628E-01  -5.495152588429E-02   0.057947436237425
 O002   O   1.009931816969E-01   4.354358061833E-01   0.056896021978425006
 O003   O  -1.627534269860E-01   2.136789405442E-01   0.054297918384875
 O004   O   3.510920646616E-01  -2.598116874907E-01   0.058013269597275
 O005   O   1.333627974677E-01  -4.496032196362E-02   0.056516663773625
 O006   O  -3.490906343680E-01   4.509506557444E-01   0.05610817853365
 O007   O   3.710164316775E-01   2.092912654339E-01   0.054461228483875
 O008   O  -1.033338929791E-01  -2.801172945248E-01   0.058895670993575
Fe009  Fe   1.155166502800E-01   2.922228960484E-01   0.0332568750888
Fe010  Fe  -3.782192237566E-01  -2.025359721149E-01   0.036837831104025
Co011  Co   3.636865101146E-01   4.677845157493E-02   0.032490255589075
Fe012  Fe  -1.320606575756E-01  -4.622067148105E-01   0.03585478199955
Fe013  Fe   1.145816191190E-01  -1.987963361893E-01   0.0339202734593
Co014  Co  -3.821102306562E-01   2.937279927937E-01   0.030722195469175
 O015   O   3.434999982295E-01  -1.153006858018E-01   0.0058013405683675
 O016   O  -1.330085758288E-01   3.723636628473E-01   0.0046659273626375
 O017   O  -4.262025187941E-01   1.373389221991E-01   0.0033520741208349996
 O018   O   9.795388636873E-02  -3.632864614066E-01   0.0066787908856225
 O019   O  -1.339581591454E-01  -1.364614122614E-01   0.0063623220624200001
 O020   O   3.810947159407E-01   3.690136345124E-01   0.0032317547561624997
 O021   O   1.214105055942E-01   1.441056974284E-01   0.002628533719555
 O022   O  -3.646707017031E-01  -3.610964450524E-01   0.009655877033735
Co023  Co  -1.505173872577E-01   4.949566595337E-02  -0.00688250908179
Fe024  Fe   3.701250937907E-01  -4.487222324289E-01  -0.0086331828551775
Fe025  Fe  -1.596415066591E-01  -2.870061923852E-01  -0.033311477528974995
Co026  Co   3.473329440553E-01   2.249926667129E-01  -0.034782003195775
Fe027  Fe   3.341270968049E-01  -1.142694440060E-01  -0.0450403556154
Fe028  Fe  -1.330209081334E-01   3.724168580030E-01  -0.044568203989
 O029   O   7.646133929833E-02  -2.004898554591E-01  -0.054319015294175
 O030   O  -3.811824736897E-01   2.731274091874E-01  -0.056315718777674995
 O031   O   3.341913601337E-01   6.159608059546E-02  -0.0572373645131
 O032   O  -1.916499123626E-01  -4.590059603193E-01  -0.052288494964599995
 O033   O  -4.274989948242E-01  -2.172806361210E-01  -0.052115058289475005
 O034   O   1.479730603197E-01   3.231956072669E-01  -0.0550302307645
 O035   O  -1.664277973347E-01   3.538640239326E-02  -0.050548631120550004
 O036   O   4.484043253985E-01  -4.177072692750E-01  -0.053388162630875
```

## (111) CFO slab 0.75 inversion spinel with H on top [octahedral] layer

```
data_CoFe2O4_111_slab_inv_Htop

_cell_length_a                      5.97660113
_cell_length_b                      10.18575093
_cell_length_c                      40.00000000
_cell_angle_alpha                   90.000000
_cell_angle_beta                    90.000000
_cell_angle_gamma                   90.338100
_symmetry_space_group_name_H-M      'P 1'
_symmetry_Int_Tables_number         1

loop_
_symmetry_equiv_pos_as_xyz
  'x, y, z'

loop_
   _atom_site_label
   _atom_site_type_symbol
   _atom_site_fract_x
   _atom_site_fract_y
   _atom_site_fract_z
 O001   O   -0.3749985679  -0.0451068623   0.0612435216
 O002   O    0.1149611326   0.4522227233   0.0628013086
 O003   O   -0.1089350417   0.2064781017   0.0586916832
 O004   O    0.3684303222  -0.2792121610   0.0626364109
 O005   O    0.1319217215  -0.0447451914   0.0619355042
 O006   O   -0.3498173107   0.4541870245   0.0639956218
 O007   O    0.3713465769   0.2136761723   0.0596931055
 O008   O   -0.1330275653  -0.2880384220   0.0610882757
Fe009   Fe   0.1319869141   0.2915514905   0.0308146393
Fe010   Fe  -0.3904689974  -0.2014978588   0.0334067707
Co011   Co   0.3741872520   0.0423790202   0.0330088444
Fe012   Fe  -0.1226744156  -0.4642570503   0.0353762943
Fe013   Fe   0.1198857668  -0.2008179735   0.0339892167
Co014   Co  -0.3704061609   0.2930030538   0.0328085106
 O015   O    0.3634311117  -0.1347695472   0.0048860209
 O016   O   -0.1146123873   0.3815911802   0.0051309345
 O017   O   -0.3982370950   0.1322459385   0.0019510004
 O018   O    0.1115178428  -0.3676250936   0.0084891072
 O019   O   -0.1402699322  -0.1403370249   0.0035447679
 O020   O    0.3673286017   0.3739464304   0.0036533418
 O021   O    0.1342977584   0.1346737515   0.0021206935
 O022   O   -0.3487854017  -0.3689026891   0.0067284259
Co023   Co  -0.1347751892   0.0473995932  -0.0090766647
Fe024   Fe   0.3747887330  -0.4496038763  -0.0084394180
Fe025   Fe  -0.1391904623  -0.2930955248  -0.0305626930
Co026   Co   0.3738364385   0.1905512227  -0.0343228464
Fe027   Fe   0.3253270492  -0.1318000933  -0.0430812341
Fe028   Fe  -0.1219333634   0.3845516532  -0.0417670639
 O029   O    0.0405935233  -0.1984013075  -0.0534894324
 O030   O   -0.3793579944   0.2873742693  -0.0526685185
 O031   O    0.3691074875   0.0382119286  -0.0542357100
 O032   O   -0.1040030418  -0.4440703401  -0.0510132509
```

```
O033   O   -0.4487967769   -0.2602988204   -0.0530483783
O034   O    0.1312418926    0.2865831378   -0.0536922010
O035   O   -0.1302193047    0.0351063804   -0.0524745385
O036   O    0.4172844093   -0.3827894175   -0.0521665845
H037   H   -0.4326724401   -0.0476754951    0.0839896879
H038   H    0.0398551362    0.4262088379    0.0834699831
H039   H   -0.1277206281    0.2336439789    0.0818525164
H040   H    0.3604038303   -0.2331877581    0.0840030666
H041   H    0.0008987821    0.0139793286    0.0627705139
H042   H   -0.4957078138    0.4976169776    0.0658331693
H043   H    0.3665764451    0.2178092185    0.0838686239
H044   H   -0.1306417714   -0.2818110050    0.0852901302
```

## (111) CFO slab 0.75 inversion spinel with H on bottom [octahedral+tetrahedral] layer

```
data_CpFe2O4_111_slab_inv_Hbot

_cell_length_a                  5.87505140
_cell_length_b                 10.34300155
_cell_length_c                 40.00000000
_cell_angle_alpha              90.000000
_cell_angle_beta               90.000000
_cell_angle_gamma              90.468400
_symmetry_space_group_name_H-M  'P 1'
_symmetry_Int_Tables_number     1

loop_
_symmetry_equiv_pos_as_xyz
   'x, y, z'

loop_
   _atom_site_label
   _atom_site_type_symbol
   _atom_site_fract_x
   _atom_site_fract_y
   _atom_site_fract_z
 O001    O   -0.4203354624   -0.0384796642    0.0545198034
 O002    O    0.0821854090    0.4390880166    0.0539213628
 O003    O   -0.1219940988    0.2339614473    0.0561118007
 O004    O    0.3859747580   -0.2693357496    0.0553560501
 O005    O    0.1522081813   -0.0340320868    0.0550557462
 O006    O   -0.3629239603    0.4690904417    0.0531546024
 O007    O    0.3737018698    0.2192726074    0.0572535569
 O008    O   -0.1548038078   -0.2951035266    0.0538382087
Fe009   Fe    0.1154391864    0.2956195980    0.0320858224
Fe010   Fe   -0.3949718103   -0.2087307263    0.0316998990
Co011   Co    0.3558027253    0.0493677335    0.0320500893
Fe012   Fe   -0.1536254577   -0.4609509162    0.0311506878
Fe013   Fe    0.1077391846   -0.2120564161    0.0330766389
Co014   Co   -0.3958989758    0.2916635156    0.0339536981
 O015    O    0.3582942854   -0.1346147649    0.0013244098
 O016    O   -0.1588120556    0.3774078200    0.0005408678
 O017    O   -0.4254549961    0.1275167415    0.0038070000
 O018    O    0.0825996579   -0.3724671092    0.0022486995
 O019    O   -0.1541638849   -0.1422009960    0.0034969272
 O020    O    0.3486752501    0.3643531333    0.0021212597
```

```
O021    O   0.1315158239   0.1399992962   0.0050755905
O022    O  -0.3760604424  -0.3694776725  -0.0018589372
Co023   Co -0.1428885357   0.0435872427  -0.0065257286
Fe024   Fe  0.3541717047  -0.4617552099  -0.0096501615
Fe025   Fe -0.1402235971  -0.2910411582  -0.0325123147
Co026   Co  0.3607974105   0.2082776595  -0.0313770952
Fe027   Fe  0.3348515694  -0.1328859798  -0.0445828446
Fe028   Fe -0.1385412317   0.3777949554  -0.0455695076
O029    O   0.0572506820  -0.1863059468  -0.0586119550
O030    O  -0.3892341822   0.2840099467  -0.0629198053
O031    O   0.3819899944   0.0393321065  -0.0621420660
O032    O  -0.1003474536  -0.4481014149  -0.0608214512
O033    O  -0.4093023455  -0.2299274919  -0.0619292171
O034    O   0.1276735397   0.2982973306  -0.0624065647
O035    O  -0.1546360960   0.0316420968  -0.0578642380
O036    O   0.3378887107  -0.4636125702  -0.0572076621
H037    H  -0.0829310862  -0.0571747368  -0.0615353316
H038    H  -0.3359864943   0.2170409692  -0.0781437207
H039    H   0.2796084960   0.0625651835  -0.0804408310
H040    H   0.0710605948  -0.4420141920  -0.0650083275
H041    H  -0.3497905067  -0.2124558480  -0.0842524409
H042    H   0.2338161651   0.3662584810  -0.0705722969
H043    H  -0.3197053395   0.0265727650  -0.0633780896
H044    H   0.4404009584  -0.3987120073  -0.0671641639
```

## (100) CFO slab 0.75 inversion spinel without H

```
data_CoFe2O4_100_slab_inv_noH

_cell_length_a                     5.93156736
_cell_length_b                     11.47283384
_cell_length_c                     40.00000000
_cell_angle_alpha                  90.000000
_cell_angle_beta                   90.000000
_cell_angle_gamma                  91.073042
_symmetry_space_group_name_H-M     'P 1'
_symmetry_Int_Tables_number        1

loop_
_symmetry_equiv_pos_as_xyz
   'x, y, z'

loop_
   _atom_site_label
   _atom_site_type_symbol
   _atom_site_fract_x
   _atom_site_fract_y
   _atom_site_fract_z
 O001    O  -0.4673132545   0.0136773403   0.0549442698
 O002    O  -0.4730561619   0.4977567478   0.0548952530
 O003    O  -0.0431704077   0.0041918670   0.0562995351
 O004    O  -0.0218526277   0.4962747851   0.0550822466
Co005   Co  -0.2607895723   0.1231670332   0.0500330276
Fe006   Fe  -0.2408533121  -0.3769720801   0.0489032443
Fe007   Fe  -0.2535459228   0.3768985466   0.0487782738
Co008   Co  -0.2356497368  -0.1218829487   0.0498141421
```

| | | | | |
|---|---|---|---|---|
| O009 | O | -0.4668033039 | 0.2440541955 | 0.0485052257 |
| O010 | O | -0.4496699451 | -0.2454994398 | 0.0510490422 |
| O011 | O | -0.0495052119 | 0.2391741830 | 0.0494303285 |
| O012 | O | -0.0158777349 | -0.2508541413 | 0.0476743380 |
| Co013 | Co | 0.2420851033 | 0.0040327389 | 0.0322519003 |
| Fe014 | Fe | 0.2523492828 | 0.4993655028 | 0.0318932459 |
| O015 | O | -0.2465213786 | -0.1274700190 | -0.0007862463 |
| O016 | O | -0.2476931073 | 0.3719578496 | -0.0007239119 |
| O017 | O | -0.2582481454 | 0.1260097367 | -0.0016833934 |
| O018 | O | -0.2516111432 | -0.3729354338 | -0.0002192143 |
| Fe019 | Fe | 0.4995640398 | -0.2465804307 | 0.0009856668 |
| Fe020 | Fe | 0.4974429416 | 0.2460145329 | -0.0016033833 |
| Fe021 | Fe | -0.0052456905 | 0.2457915344 | -0.0016003005 |
| Co022 | Co | 0.0061004022 | -0.2485183363 | 0.0001982030 |
| O023 | O | 0.2601041573 | -0.1286786279 | 0.0010349877 |
| O024 | O | 0.2496749524 | 0.3706885088 | 0.0011126840 |
| O025 | O | 0.2413734549 | 0.1301553427 | -0.0010211145 |
| O026 | O | 0.2551417003 | -0.3724512239 | 0.0018417706 |
| Fe027 | Fe | -0.2570384654 | -0.0031859283 | -0.0324156097 |
| Fe028 | Fe | -0.2440244976 | -0.4980853447 | -0.0319148609 |
| O029 | O | 0.0316102431 | -0.2468548470 | -0.0474089879 |
| O030 | O | 0.0276092346 | 0.2525003550 | -0.0500521990 |
| O031 | O | 0.4657203778 | -0.2537909148 | -0.0488784122 |
| O032 | O | 0.4648969118 | 0.2498675645 | -0.0502210983 |
| Co033 | Co | 0.2439287896 | -0.1227842913 | -0.0484757652 |
| Fe034 | Fe | 0.2541165428 | 0.3788040587 | -0.0493745357 |
| Fe035 | Fe | 0.2400188164 | 0.1220175912 | -0.0506462714 |
| Co036 | Co | 0.2599454616 | -0.3757265214 | -0.0483036098 |
| O037 | O | 0.0133991952 | 0.0002413121 | -0.0553729674 |
| O038 | O | 0.0341525124 | -0.4941861198 | -0.0542615955 |
| O039 | O | 0.4652089429 | -0.0043808855 | -0.0553318388 |
| O040 | O | 0.4845807518 | 0.4984042930 | -0.0543420321 |

### (100) CFO slab 0.75 inversion spinel with H

**data_CoFe2O4_100_slab_inv_H**

**_cell_length_a 5.91035621**
**_cell_length_b 11.44890558**
**_cell_length_c 40**
**_cell_angle_alpha 90.000000**
**_cell_angle_beta 90.000000**
**_cell_angle_gamma 91.615476**
**_symmetry_space_group_name_H-M     'P 1'**
**_symmetry_Int_Tables_number        1**

**loop_**
**_symmetry_equiv_pos_as_xyz**
   **'x, y, z'**

**loop_**
**_atom_site_label**
**_atom_site_type_symbol**
**_atom_site_fract_x**
**_atom_site_fract_y**
**_atom_site_fract_z**

```
 O001   O  -4.742149551914E-01   1.132012080883E-02   0.053501210578799994
 O002   O  -4.680753152845E-01   4.980632339307E-01   0.054310376787750005
 O003   O  -4.977973871053E-02   4.300382932541E-03   0.05515424603932499
 O004   O  -1.942353184620E-02   4.965888296374E-01   0.054003019866699996
Co005  Co  -2.651434176083E-01   1.234666027597E-01   0.051272942396525
Fe006  Fe  -2.359004271697E-01  -3.760831257782E-01   0.050742106008675
Fe007  Fe  -2.512356969210E-01   3.756478491004E-01   0.050767790801325
Co008  Co  -2.357335252080E-01  -1.232794763075E-01   0.051313013040474996
 O009   O  -4.687971924850E-01   2.416115381706E-01   0.0486939132909
 O010   O  -4.514389720718E-01  -2.448512808551E-01   0.051141738552425
 O011   O  -4.785103660278E-02   2.400704839089E-01   0.0492473766756
 O012   O  -1.317313746839E-02  -2.499509710185E-01   0.048110142162175004
Co013  Co   2.353990693043E-01   3.826769599312E-03   0.030453087531475
Fe014  Fe   2.563767737395E-01   4.996300984375E-01   0.030797169955975
 O015   O  -2.449378485156E-01  -1.277793874222E-01  -0.000172266437497025
 O016   O  -2.463949445761E-01   3.710273299173E-01   7.5760374435525e-05
 O017   O  -2.579475573439E-01   1.273866372159E-01  -0.0018472447869225
 O018   O  -2.514406701576E-01  -3.707516576143E-01   0.0006642967239095
Fe019  Fe  -4.993863627174E-01  -2.467543041865E-01   0.00118703711723
Fe020  Fe   4.990818221322E-01   2.460015852627E-01  -0.0017012842953329998
Fe021  Fe  -3.640707661210E-03   2.472719283805E-01  -0.0009998508980374999
Co022  Co   7.347670371634E-03  -2.473515886670E-01   0.00026312669883149997
 O023   O   2.621342646947E-01  -1.290747703335E-01   0.00061932479148225
 O024   O   2.543001106787E-01   3.698194694067E-01   0.0007928393746595
 O025   O   2.404030541556E-01   1.310344373676E-01  -0.0017286113989642498
 O026   O   2.543818863922E-01  -3.706072194091E-01   0.0017972487558975
Fe027  Fe  -2.561947030770E-01  -2.794744225178E-03  -0.03132267677375
Fe028  Fe  -2.428773187746E-01  -4.976793353637E-01  -0.030150071133724997
 O029   O   2.988143515236E-02  -2.447231270921E-01  -0.047507432703375
 O030   O   2.641633914260E-02   2.563000024299E-01  -0.049983319532675
 O031   O   4.697732820099E-01  -2.554046782295E-01  -0.04896054506975
 O032   O   4.665243134517E-01   2.472694253083E-01  -0.050531292184700004
Co033  Co   2.483397725247E-01  -1.231357603197E-01  -0.050290251846925
Fe034  Fe   2.570883941888E-01   3.784253416562E-01  -0.051096340021575
Fe035  Fe   2.365935333678E-01   1.239909671716E-01  -0.052315840403650005
Co036  Co   2.587170293207E-01  -3.742293095763E-01  -0.050007564345175004
 O037   O   1.561248470401E-02   3.006759138515E-03  -0.054402465299424996
 O038   O   3.620494243651E-02  -4.917666634845E-01  -0.052846409155925
 O039   O   4.656026206490E-01  -6.263787827588E-03  -0.0546695785001
 O040   O   4.858346460725E-01   4.967835265853E-01  -0.052950504634350005
 H041   H  -2.700428793881E-01   1.360931269825E-01   0.08701727728325001
 H042   H  -2.292079805065E-01  -3.551791321684E-01   0.0895294961979
 H043   H  -2.476801177650E-01   3.530112205665E-01   0.089403273132275
 H044   H  -2.502783617492E-01  -1.417584598446E-01   0.088080296117425
 H045   H   2.151137316257E-01  -1.439960597455E-01  -0.086917885363125
 H046   H   2.432363187437E-01   3.610789366792E-01  -0.08975087829155
 H047   H   2.472232204528E-01   1.427172992649E-01  -0.0909604290744
 H048   H   2.769349448890E-01  -3.585995893866E-01  -0.0870068254091
```

## (111) NFO slab 1.0 inversion spinel (without H)

`data_NiFe2O4_111_slab_inv_noH`

| | |
|---|---|
| `_cell_length_a` | 5.81681872 |
| `_cell_length_b` | 5.76025247 |
| `_cell_length_c` | 40.00000000 |

```
_cell_angle_alpha                          90.000000
_cell_angle_beta                           90.000000
_cell_angle_gamma                          119.540191
_symmetry_space_group_name_H-M             'P 1'
_symmetry_Int_Tables_number                1

loop_
_symmetry_equiv_pos_as_xyz
   'x, y, z'

loop_
   _atom_site_label
   _atom_site_type_symbol
   _atom_site_fract_x
   _atom_site_fract_y
   _atom_site_fract_z
 O001    O   -0.3915847788   -0.3209328539    0.0556394240
 O002    O    0.0469960339    0.1817210598    0.0571396376
 O003    O   -0.3736391463    0.1711502037    0.0564238540
 O004    O    0.0514215740   -0.3589235264    0.0540748179
 Ni005   Ni  -0.0923408998   -0.1747978444    0.0309458644
 Fe006   Fe  -0.0846257901    0.3337650582    0.0322936121
 Ni007   Ni   0.4199680082   -0.1735449176    0.0327080213
 O008    O   -0.2307536451    0.0113073549    0.0045081325
 O009    O   -0.2382259805    0.4815840269    0.0008160062
 O010    O    0.2321519297   -0.4990138688    0.0020770022
 O011    O    0.2117979236   -0.0428635230    0.0049902644
 Fe012   Fe   0.3968759573    0.3042861143   -0.0115618394
 Fe013   Fe   0.0792582753   -0.3416110126   -0.0326900283
 Fe014   Fe  -0.2270651597    0.0359282645   -0.0448032119
 O015    O   -0.0125242047    0.3626895142   -0.0577366399
 O016    O    0.4153266256   -0.0564404499   -0.0535244402
 O017    O   -0.1232289887   -0.2004732301   -0.0581393240
 O018    O    0.3672326900    0.1722162891   -0.0563081962
```

**(111) NFO slab 1.0 inversion spinel with H on top [octahedral] layer**

```
data_NiFe2O4_111_slab_inv_Htop

_cell_length_a                             5.71430895
_cell_length_b                             5.91779889
_cell_length_c                             60.00000000
_cell_angle_alpha                          90.000000
_cell_angle_beta                           90.000000
_cell_angle_gamma                          118.903400
_symmetry_space_group_name_H-M             'P 1'
_symmetry_Int_Tables_number                1

loop_
_symmetry_equiv_pos_as_xyz
   'x, y, z'

loop_
   _atom_site_label
   _atom_site_type_symbol
   _atom_site_fract_x
```

```
    _atom_site_fract_y
    _atom_site_fract_z
 O001    O   -0.3850781099  -0.3033233884   0.0396413144
 O002    O    0.0606172409   0.1586571370   0.0411857427
 O003    O   -0.4250434687   0.1695393458   0.0402570910
 O004    O    0.0648678020  -0.3323853349   0.0383355468
 Ni005   Ni  -0.0858792073  -0.1607327695   0.0211350435
 Fe006   Fe  -0.0969160044   0.3198497773   0.0214652710
 Ni007   Ni   0.4158922978  -0.1616471180   0.0218122782
 O008    O   -0.2412056935  -0.0031536383   0.0036359955
 O009    O   -0.2370752382   0.4946206634   0.0007112785
 O010    O    0.2272724526  -0.4889915104   0.0022952727
 O011    O    0.2161042118  -0.0249910541   0.0031073589
 Fe012   Fe   0.3938105340   0.3185030154  -0.0079401269
 Fe013   Fe   0.0711741813  -0.3580629515  -0.0212985662
 Fe014   Fe  -0.2601916941  -0.0409176727  -0.0289870526
 O015    O   -0.0762721444   0.3014870717  -0.0370059910
 O016    O    0.3815352131  -0.1963058730  -0.0342809452
 O017    O   -0.0976707698  -0.2351774502  -0.0363165665
 O018    O    0.3068222242   0.3043842081  -0.0377568180
 H019    H   -0.3153379654  -0.1916830778   0.0527432233
 H020    H    0.2569987256   0.2310045413   0.0415118961
 H021    H   -0.3757440177   0.1526162848   0.0554715719
 H022    H    0.0402359602  -0.3199902208   0.0542059502
```

**(111) NFO slab 1.0 inversion spinel with H on bottom [octahedral+tetrahedral] layer**

```
data_NiFe2O4_111_slab_inv_Hbot

_cell_length_a                      5.85757200
_cell_length_b                      5.93817822
_cell_length_c                      60.00000000
_cell_angle_alpha                   90.000000
_cell_angle_beta                    90.000000
_cell_angle_gamma                   120.003400
_symmetry_space_group_name_H-M      'P 1'
_symmetry_Int_Tables_number         1

loop_
_symmetry_equiv_pos_as_xyz
   'x, y, z'

loop_
    _atom_site_label
    _atom_site_type_symbol
    _atom_site_fract_x
    _atom_site_fract_y
    _atom_site_fract_z
 O001    O   -0.4195670596  -0.3489330392   0.0343365609
 O002    O    0.0267706710   0.1339226075   0.0365626260
 O003    O   -0.3810530839   0.1520139288   0.0365640733
 O004    O    0.0413262613  -0.3428210363   0.0344972708
 Ni005   Ni  -0.1092969014  -0.1882725417   0.0185680301
 Fe006   Fe  -0.0974217676   0.3241504311   0.0221243989
 Ni007   Ni   0.4006565542  -0.1851310501   0.0205574246
 O008    O   -0.2655048398  -0.0324461342   0.0002050621
```

```
O009    O   -0.2586965775    0.4648047716   -0.0024331502
O010    O    0.2071929373    0.4806139178    0.0002061902
O011    O    0.1962848608   -0.0584666878   -0.0008992569
Fe012   Fe   0.3806092843    0.2964043865   -0.0072302654
Fe013   Fe   0.0570121179   -0.3723989005   -0.0212402927
Fe014   Fe  -0.2746740801   -0.0261420051   -0.0300420597
O015    O   -0.0529422433    0.3152107051   -0.0414336788
O016    O    0.3761647099   -0.1509433657   -0.0407368859
O017    O   -0.1448781936   -0.2442542370   -0.0419676297
O018    O    0.3733098986    0.2864022333   -0.0397475934
H019    H    0.1224854950    0.3236087144   -0.0444288335
H020    H    0.3695997471    0.0167444802   -0.0429516747
H021    H   -0.0542068176   -0.1916518267   -0.0563053873
H022    H   -0.4590565118    0.4325613556   -0.0453271009
```

## (100) NFO slab 1.0 inversion spinel without H

```
data_NiFe2O4_100_slab_inv_noH

_cell_length_a                          5.61233711
_cell_length_b                          5.86083030
_cell_length_c                         40.00000000
_cell_angle_alpha                      90.000000
_cell_angle_beta                       90.000000
_cell_angle_gamma                      89.339300
_symmetry_space_group_name_H-M         'P 1'
_symmetry_Int_Tables_number            1

loop_
_symmetry_equiv_pos_as_xyz
   'x, y, z'

loop_
   _atom_site_label
   _atom_site_type_symbol
   _atom_site_fract_x
   _atom_site_fract_y
   _atom_site_fract_z
O001    O    0.0065513437   -0.4689123789    0.0551860036
Fe002   Fe   0.2452277563   -0.2538525905    0.0496265195
O003    O    0.4992824748   -0.4536012716    0.0519077035
Ni004   Ni  -0.2469173152   -0.2524008022    0.0503908878
O005    O    0.0036049359   -0.0370722159    0.0556865994
O006    O    0.4943682700   -0.0506128922    0.0508804906
Fe007   Fe   0.0003504653    0.2463015255    0.0309572981
O008    O   -0.2543640302   -0.2497913162   -0.0001965906
O009    O    0.2487925143   -0.2466894754    0.0005854946
Fe010   Fe   0.4991512364   -0.0054292686   -0.0019258577
Ni011   Ni   0.4974912508    0.4980956044   -0.0002587242
O012    O   -0.2554248559    0.2412995918    0.0014094464
O013    O    0.2538928386    0.2434364425    0.0017808975
Fe014   Fe   0.0031745546   -0.2472079126   -0.0319221331
O015    O    0.4875384443    0.0351561435   -0.0494648386
Fe016   Fe  -0.2454541075    0.2579172059   -0.0493517697
O017    O    0.0000017427    0.0281111279   -0.0553310090
Ni018   Ni   0.2435724864    0.2499024240   -0.0502233878
```

```
 O019   O   0.4820059246   0.4595917618  -0.0502584251
 O020   O   0.0014913328   0.4758335109  -0.0548564641
```

**(100) NFO slab 1.0 inversion spinel with H**

```
data_NiFe2O4_100_slab_inv_H

_cell_length_a 5.59038549
_cell_length_b 5.83789352
_cell_length_c 40
_cell_angle_alpha 90.000000
_cell_angle_beta 90.000000
_cell_angle_gamma 89.369600
_symmetry_space_group_name_H-M          'P 1'
_symmetry_Int_Tables_number             1

loop_
_symmetry_equiv_pos_as_xyz
   'x, y, z'

loop_
_atom_site_label
_atom_site_type_symbol
_atom_site_fract_x
_atom_site_fract_y
_atom_site_fract_z
  O001   O    6.357450507867E-03 -4.667698494075E-01  0.054606660306275
 Fe002  Fe    2.482166978516E-01 -2.544056208954E-01  0.0519209754397
  O003   O    4.999931707081E-01 -4.564374665385E-01  0.051777627065775
 Ni004  Ni   -2.491758389287E-01 -2.525758381136E-01  0.0521212519231
  O005   O    2.539490070244E-03 -3.927895267340E-02  0.055306816134125004
  O006   O    4.949220301840E-01 -4.893795544527E-02  0.051481352700475005
 Fe007  Fe   -2.194038769692E-03  2.455679266035E-01  0.0299260801568
  O008   O   -2.551297705461E-01 -2.485131617547E-01  0.00015481522464337502
  O009   O    2.519335129598E-01 -2.458615658015E-01  0.0013152668641502499
 Fe010  Fe    4.986009083242E-01 -6.485487062857E-03 -0.0023256641217450002
 Ni011  Ni    4.977270925916E-01  4.976443048096E-01 -0.00043816457860199997
  O012   O   -2.573399203155E-01  2.398686944685E-01  0.0006302352125067501
  O013   O    2.566760997620E-01  2.425524207160E-01  0.002341316176053
 Fe014  Fe    5.398155567061E-03 -2.456050344456E-01 -0.030629953083875
  O015   O    4.832581603410E-01  3.261915249029E-02 -0.050165093913500006
 Fe016  Fe   -2.465378865988E-01  2.582995673024E-01 -0.051794558024475
  O017   O    8.242774681382E-04  3.139001667594E-02 -0.054756984494875004
 Ni018  Ni    2.460837648445E-01  2.508922808854E-01 -0.051938554234949995
  O019   O    4.841781305546E-01  4.608551423184E-01 -0.050651337141899996
  O020   O    4.557261077439E-03  4.762937923788E-01 -0.053917587316325
  H021   H    2.756045107173E-01 -2.518846880138E-01  0.0897110870841
  H022   H   -2.608616542299E-01 -2.535273556286E-01  0.087114119925625
  H023   H   -2.901991931262E-01  2.536120479247E-01 -0.09010633422685001
  H024   H    2.653336718001E-01  2.523290805911E-01 -0.086618982201025
```

**References:**

(1) Sanchez-Lievanos, K. R.; Tariq, M.; Brennessel, W. W.; Knowles, K. E. Heterometallic Trinuclear Oxo-Centered Clusters as Single-Source Precursors for Synthesis of Stoichiometric Monodisperse Transition Metal Ferrite Nanocrystals. *Dalton Trans.* **2020**, *49* (45), 16348–16358.

(2) Choi, J.; Suryanto, B. H. R.; Wang, D.; Du, H.-L.; Hodgetts, R. Y.; Ferrero Vallana, F. M.; MacFarlane, D. R.; Simonov, A. N. Identification and Elimination of False Positives in Electrochemical Nitrogen Reduction Studies. *Nat. Commun.* **2020**, *11* (1), 5546.

(3) Dovesi, R.; Erba, A.; Orlando, R.; Zicovich-Wilson, C. M.; Civalleri, B.; Maschio, L.; Rérat, M.; Casassa, S.; Baima, J.; Salustro, S.; Kirtman, B. Quantum-Mechanical Condensed Matter Simulations with CRYSTAL. *WIREs Comput. Mol. Sci.* **2018**, *8* (4), e1360.

(4) Heyd, J.; Scuseria, G. E.; Ernzerhof, M. Hybrid Functionals Based on a Screened Coulomb Potential. *J. Chem. Phys.* **2003**, *118* (18), 8207–8215.

(5) Grimme, S.; Antony, J.; Ehrlich, S.; Krieg, H. A Consistent and Accurate Ab Initio Parametrization of Density Functional Dispersion Correction (DFT-D) for the 94 Elements H-Pu. *J. Chem. Phys.* **2010**, *132* (15), 154104.

(6) Peintinger, M. F.; Oliveira, D. V.; Bredow, T. Consistent Gaussian Basis Sets of Triple-Zeta Valence with Polarization Quality for Solid-State Calculations. *J. Comput. Chem.* **2013**, *34* (6), 451–459.

(7) Sanchez-Lievanos, K. R.; Knowles, K. E. Controlling Cation Distribution and Morphology in Colloidal Zinc Ferrite Nanocrystals. *Chem. Mater.* **2022**, *34* (16), 7446–7459.